\documentclass[useAMS,usenatbib]{mn2e}

\usepackage{graphicx}

\title[Origin of ISM turbulence]
{The onset of large scale turbulence in the interstellar medium of spiral galaxies}
\author[D. Falceta-Gon\c{c}alves et al.]{D. Falceta-Gon\c{c}alves$^{1,2}$\thanks{E-mail:dfalceta@usp.br}; I. Bonnell$^{1}$; G. Kowal$^{2}$; J. R. D. L\'epine$^{3}$; C. A. S. Braga$^{3}$
\\$^{1}$SUPA, School of Physics \& Astronomy, University of St Andrews, North Haugh, St Andrews, Fife KY16 9SS, UK 
\\$^{2}$Escola de Artes, 
Ci\^encias e Humanidades, 
Universidade de S\~ao Paulo, Rua Arlindo Bettio 1000, CEP 03828-000,
S\~ao Paulo, Brazil
\\$^{3}$Instituto de Astronomia, Geofísica e Ciências Atmosféricas, Universidade de São Paulo, Rua do Matão 1226,  CEP: 05508-090, S\~ao Paulo, Brazil}

\begin{document}

\date{}

\pagerange{\pageref{firstpage}--\pageref{lastpage}} \pubyear{2014}

\maketitle

\label{firstpage}

\begin{abstract} 
 
Turbulence is ubiquitous in the interstellar medium (ISM) of the Milky Way and other spiral galaxies. The energy source for this turbulence has been much debated with many possible origins proposed. The universality of turbulence, its reported large-scale driving, and that it occurs also in starless molecular clouds, challenges models invoking any stellar source. A more general process is needed to explain the observations. In this work we study the role of galactic spiral arms. This is accomplished by means of three-dimensional hydrodynamical simulations which follow the dynamical evolution of interstellar diffuse clouds ($\sim 100$cm$^{-3}$) interacting with the gravitational potential field of the spiral pattern. We find that the tidal effects of the arm's potential on the cloud result in internal vorticity, fragmentation and hydrodynamical instabilities.  
The triggered turbulence result in large-scale driving, on sizes of the ISM inhomogeneities, i.e. as large as $\sim 100$pc, and  efficiencies in converting potential energy into turbulence in the range $\sim$ 10 to 25 \% per arm crossing. This efficiency is much higher than those found in previous models. The statistics of the turbulence in our simulations are strikingly similar to the observed power spectrum and Larson scaling relations of molecular clouds and the general ISM. The dependency found from different models indicate that the ISM turbulence is mainly related to local spiral arm properties, such as its mass density and width. This correlation seems in agreement with recent high angular resolution observations of spiral galaxies, e.g. M51 and M33.
\end{abstract}

\begin{keywords} 
turbulence - 
stars: formation -
ISM: general, clouds, kinematics and dynamics -
methods: numerical
\end{keywords}
      
\section{Introduction}

Interstellar turbulence has been inferred observationally since the 1950's \citep{von51}, 
based on the spatial distribution of the interstellar matter over the plane of sky. Its complex/filamentary structure resembled those also observed from chaotic motions of turbulent flows. Velocity 
dispersions have also been measured by that time \citep{vonH51}, revealing the 
supersonic nature of the turbulent motions. For the years that followed, the view of 
a turbulence dominated interstellar medium (ISM) became much clearer 
\cite[see reviews by][and references therein]{elm04, mac04, hen12, fal14}.

H and CO surveys from molecular clouds reveals an universal scaling 
relation of the measured line-widths, with $\sigma_v \propto l^\alpha$ \citep{lar81}, over lenghtscales 
ranging from 0.01pc up to tens of parsecs \citep{hey04, gol08, hey09, liu12, poidevin13}. The observed data is well fitted by a power-law with an $\alpha \simeq 0.5$ slope. Such an universal slope for the turbulence in the ISM is a stricking result. Exceptions to this universal behavior arise naturally at high density collapsing cores as gravitational effects may dominate \citep{yos10,hey12} though.
Density fluctuations probed by scintillation of backround radiation, as well as rotation measure of intrinsic interstellar emission \citep{arm95,min96}, reveals a turbulent picture of diffuse ISM as well. The observed data indicate a single power-law for the whole interstellar turbulence up to hundreds of parsecs in lenghtscale. Therefore the current paradigm points towards an universal origin for the turbulence in the Galaxy. 
What would then be large scale driving source of turbulence in our Galaxy?

Large star forming regions in our Galaxy, such as Carina and Orion nebulae, induced theorists to assume a maximum role of stellar feedback. Winds, ionization fronts and, most of all, Supernovae (SNe), have been claimed 
as dominant sources for the kinetic energy of the ISM \citep[e.g.][and others]{mac04,gres08,hill12}.
Despite providing similar energy injection rate compared to the estimates for the turbulent ISM, 
stellar feedback is not universal, and happens at much smaller scales. As pointed by \citet{hei06}, 
stellar feedback acts locally, and only after the first stars are formed in the
cloud. Also, numerical simulations 
reveal that supperbubbles tend to release most of its energy perpendicular to 
the galactic disk \citep[e.g.][]{melioli09,hen10}, reducing their efficiency in maintaining the disk gas turbulent. Moreover, it is hard to relate stellar feedback to the turbulent motions of 
quiescent molecular clouds. These objects show little star formation and are too dense and cold for 
external sources, such as a blast wave, to have any effect in driving internal turbulent motions. Most 
of these objects present similar internal kinematics though \citep[see e.g.][]{will00,poidevin13}. 

Self-gravity has also been previously proposed as major driver of internal motions, on the lengthscales of clouds \citep[e.g.][]{vs2008}, as well as globally in the galactic disk \citep[e.g.][]{wada02,kim06,wada07,age09}. 
These later works rely on the fact that radiative cooling of the insterstellar gas results in the formation of regions in the disk that become gravitationally unstable, i.e. Toomre $Q<1$. The dynamical evolution of the disk after collapse was followed and, in general, the motions observed have been interpreted as turbulence. \citet{wada02} showed that the collapse first drives motions at smaller scales, which then grow to larger scales, in agreement with a type of inverse-cascade of the fluctuations. The power spectrum of their simulations presented inertial ranges with slopes of $\simeq -0.8$, flatter than observed. Since the efficiency of such mechanism depends on the initial temperature, as $Q_c \propto c_s$, the gravitational collapse generally starts once the gas is cool and mostly neutral, or molecular. The low temperatures therefore result in low velocity turbulence. Another potential problem with this mechanism is that the motions driven by self-gravity are largely coherent and hence are less likely to drive the chaotic motions inherent in turbulence.
Furthermore, as turbulence does not reside solely in molecular clouds that are self-gravitating, a non-self-gravity origin for turbulence is required.

Converging flows have been considered as one of the main mechanisms for the formation of molecular 
clouds \citep{aud05,hei06,hen07,ban09,hei11}, and could also be the cause of their turbulence. Strong shocks 
combined to efficient cooling of the downstream gas result in very dense and cold thin 
layers. Applied to the ISM, this picture may be understood as the origin of dense and cold 
structures, like the cold ISM, or even molecular clouds. These dense layers may become 
unstable to the nonlinear thin layer instability (NTLI)\citep{vish94}, resulting in 
their fragmentation and a complex velocity distribution. Converging flows, combined to 
the NTLI, would then not only be able to explain the formation of molecular clouds but also the 
internal turbulence. \citet{vs2006} showed by means of numerical simulations 
that a thin cold sheet, reminiscent of those observed
by Heiles \& Troland (2003) and Heiles (2004) in our Galaxy, can be formed at
the junction of the two converging flows. The velocity dispersion observed (attributed as 
''turbulence" by the authors) is credited to the NTLI. 
However, as shown by \citet{hei06}, the complex velocity fields observed in numerical 
simulations of converging flows is mostly due to the combined dynamics of the individual condensations 
rather than proper turbulent motions within these clouds. The energy source that generates 
the cloud-like structures cannot be the same that drives its internal turbulence. Also, shocks 
are not efficient in providing kinetic energy to the turbulent motions of the gas.
\citet{hei06} find $<5\%$ efficiency in coverting the large scale kinetic energy of the flows
 into turbulent components in the dense layer. Most of this energy is actually lost by thermal 
 radiation. Although coverging flows may be the dominant process for the formation of dense structures in the ISM, e.g. molecular clouds,  a different mechanism is responsible for the origin of their turbulence.

Since the early work of \citet{rob69}, the origin of dense clouds have also been associated to the 
interaction of interstellar gas with the gravitational potential of the spiral pattern of disk 
galaxies. The first spiral model of \citet{lin64} considered these as propagating waves, based on global distributions of stars and gas taken altogether. They showed that the non-axisymmetric disturbances can propagate in a constant shape so that they always look like a spiral arm. In a different approach, \citet{kalnajs73} analyzed the orbits of the stars in a galactic disk. He  found that, in a given frame of reference rotating with angular velocity $\Omega_P$, it is possible to construct a sequence of closed stellar orbits of increasing radii that produce enhanced stellar densities where these orbits are closer to each other. It was shown that the enhanced density of orbits is of a spiral-like shape. Since the orbits are closed, they repeat themselves after each revolution, and therefore, produce long-lived spirals. Several later works \citep[e.g.][]{jun13, pich03} showed that it is possible to obtain self-consistent solutions in this scenario. A spiral-shaped perturbation in a pre-existing axisymmetric potential modified the stellar orbits, which evolve into a new perturbation of the total potential. The solution for this self-consistent model is long-lived spirals. However, it has been also argued, based on N-body simulations, that arms formed from the crowding of stellar orbits are transient \citep[][]{sellwood84,carl85,elme93,bott03,baba09,fujii11}. See the review of \citet{dobbs14}, for a discussion on this subject. \citet{scar12} argued that the breaks in the metallicity gradient seen in spiral galaxies would not exist if the arms were short-lived. Finally, the recent analysis of spatial distributions of stars, gas and dust of the Milky Way, based on several tracers, agrees better with a long-lived density-wave theory of a 4-armed galaxy \citep[see][for details]{vallee14b}.

Recent studies \citep[e.g.][]{kim06,bonnell06, dobbs08,bonnell13} have employed numerical 
simulations to study the dynamics of the interstellar gas as it passes through a galactic spiral shock. 
These revealed that spiral shocks, associated to thermal instabilities, naturally 
give rise to a cold gas phase in the arms that develops into molecular clouds and star forming regions. 
In spite of stellar feedback models, cloud-arm interactions are ``more" universal in the sense that 
they should occur all over the galactic disk.
As the interstellar gas flows into the arms the shock fronts are in general non-steady and 
may suffer local instabilities such as the NLTI and Kelvin-Helmholtz instability (KHI). This unstable 
region may drive turbulence-like perturbations \citep{wada04,kimkim06,kkk14}. 
However, the perturbations arise away of the dense regions that evolve to molecular phase. 
The main role of the instabilities would be to drive an inhomogeneous ISM instead.

The newly born clouds may then leave the arms, thanks to the combined effect of 
centrifugal acceleration (for an observer in the reference frame of the cloud), as the cloud interacts with the arms potential, and sheared velocities 
of the gas due to the rotation of the Galaxy. Dense gas streaming out of arms is observed in spiral 
galaxies, identified as ''spurs" \citep{dobbs06}. 
At the interarm region these may be dissipated, by external heating and ionization, or survive and 
interact with subsequent spiral patterns (see Fig.\ref{fig1}). Cloud-arm interactions could then 
be an interesting alternative. 

Many decades ago \citet{wood76} provided two-dimensional numerical studies of the 
interaction of a cloud and the spiral shock, proposing that such interaction 
triggers star formation as the cloud implodes. The shocked gas should cool at timescales 
shorter than the dynamical time, resulting in a dense and cool cloud that then fragments and 
forms stars. That author also showed that sheared 
motions of the cloud and surrounding gas drives KHI that can excite 
local turbulence. More recently, \citet{bonnell06} provided a number of SPH simulations 
taking into account pre-shock interstellar medium clumpiness, and found that a spread in the 
velocity distribution of the gas has also been observed. However 
the velocity dispersion observed was identified by the authors as due to the random mass 
loading of clumps at the spiral shock, and not proper turbulence.

If focused on the shock only the mechanism of cloud-arm interaction decribed above can 
be understood as a variation of the converging flows model. 
Therefore both models suffer from the 
limitations such as energy transfer efficiency, and driving scales for the turbulence. 
However, other facet of the cloud-arm interaction has not been fully addressed yet on 
the problem of turbulence triggering: the effects of the gravitational potential of the arm on 
the cloud itself. Up to now most models have focused on the interaction of the cloud with the gas component 
of the arm. These have neglected the tidal effects that can drive internal motions, possibly more 
efficiently than the shocks. Also, as discussed further in the paper, most of the 
previous numerical simulations of the gas content of the galactic disk 
make use of sinusoidal profiles for the arm potential. Here we have chosen 
more realistic distributions obtained from self-consistent analysis of stellar 
dynamics, which result in exponential profile. 

In this work we revisit the problem of cloud-arm interactions, and 
provide a systematic study of the interaction of clouds and the gravitational 
potential of spiral pattern aiming at the onset of turbulent motions. This study is accomplished by means of full three-dimensional hydrodynamical simulation, using a grid-based Godunov scheme.  
The problem and numerical setup are described in Section 2. 
Main results from simulations are provided in Section 3. In Section 4, we
discuss the results obtained comparing them with previous works, and by providing an analytical 
toy model to the problem, followed by the main Conclusions of this work.

\section{Governing equations and model setup}

\begin{figure*}
{\centering
 \includegraphics[width=12cm]{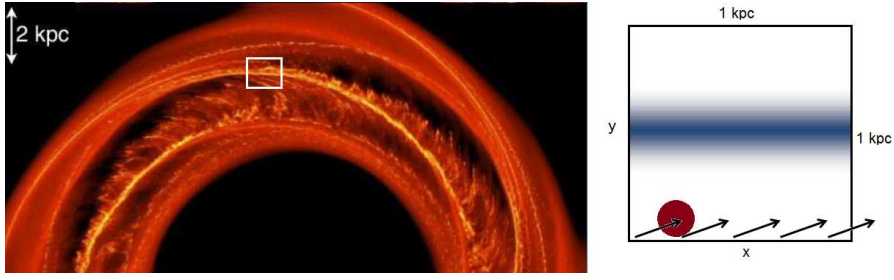}
}
 \caption{Left: global galactic simulation, obtained from \citet{bonnell13}, showing the growth of inhomogeneities (dense clouds) that move through the spiral pattern. The numerical domain of the simulations in this work is reduced, which dimensions are indicated as the white box. Right: scheme of the inital setup of the numerical simulations, indicating the initial spherical cloud moving with the ambient flow (arrows) into the deepest region of the potential well of the spiral arm depicted in grey. }
\label{fig1}
\end{figure*}

The dynamical evolution of the interstellar gas, as it interacts with the 
gravitational potential of the galactic arm, is determined by the full 
set of hydrodynamical equations, solved in the conservative form as: 

\begin{equation}
\partial_t\mbox{\bf U} + \nabla \cdot\mbox{\bf F}(\mbox{\bf U}) = f(\mbox{\bf U}),
\label{eq:genform}
\end{equation}

\noindent
where $f({\bf U})$ is the source term, ${\bf U}$ is the vector of conserved variables:

\begin{equation}
{\bf U} = \Bigl[  \rho, \rho{\bf v}, \left(\frac{1}{\gamma-1} p + \frac{1}{2}\rho v^2 \right)\Bigr]^T,
\end{equation}

\noindent
and ${\bf F}$ is the flux tensor:

\begin{eqnarray}
{\bf F} & = & \Bigl[ \rho {\bf v}, \rho {\bf v}{\bf v}+ p{\bf I}, \left(\frac{\gamma}{\gamma -1}p + \frac{1}{2}\rho v^2 \right){\bf v}\Bigr]^{T}
\label{eq:flux}
\end{eqnarray}

\noindent
where $\rho$ is the gas mass density, ${\bf I}$ the identity matrix, ${\bf v}$ the fluid velocity,
 $p$ the thermal pressure, $\gamma$ the adiabatic polytropic index, and $f$ corresponds to 
source terms for the given conserved variable $U$. The set of equations is 
closed by calculating the radiative cooling as a source term for the energy 
equation, as follows:

\begin{equation}
\frac{\partial p}{\partial t} = \frac{1}{(1-\gamma)} n^2 \Lambda(T),
\end{equation}

\noindent
where $n$ is the number density and $\Lambda(T)$ is the cooling function, which is 
obtained through an interpolation method of the electron cooling efficiency 
table for an optically thin gas.  The cooling function used was 
obtained from \citet{smith08}, for which emission lines 
from the main atoms and molecules (e.g. CO) are obtained at a temperature 
range of $T=10-10^8$K and gas densities up to $10^{12}$cm$^{-3}$, assuming solar
metallicity $Z=1Z_\odot$. 

The set of equations is solved using the GODUNOV 
code\footnote{http://amuncode.org} \citep[see][]{kowal10, fal10a, fal10b, fal10c, kowal11a, kowal11b, 
fal11, ruiz13, rei14, fm14}. The spatial reconstruction is obtained by means of the
$5^{th}$ order monotonicity-preserving (MP) method \citep{he11}, with 
approximate HLLC Riemann solver \citep{mignone06}. The time 
integration is performed with the use of a $3^{rd}$ order
four-stage explicit optimal Strong Stability Preserving Runge-Kutta SSPRK(4,3)
method \citep{ruuth06}. 

The system is set in the reference frame of the spiral arm, therefore 
the non-inertial terms (centrifugal and Coriolis) are taken as 
external source terms. The gravitational potential of the galaxy ($\Phi$) is also set as an external 
source term. self-gravity of the interstellar gas has been neglected in this work, as well as magnetic 
fields.
Therefore, the external source term in the momentum equation is given by:

\begin{equation}
{\bf f}(\rho {\bf v}) = -{\bf \Omega}_p \times \left({\bf \Omega}_p \times {\bf r}\right) - 2 {\bf \Omega}_p\times {\bf v}  -{\bf \nabla} \Phi ,
\end{equation}

\noindent
where we assume $\Omega_p$ is the angular velocity of the 
spiral pattern of the Galaxy. \citet{ger11} presented a compilation of 
estimates of $\Omega_p$ based on different methods, obtaining values in the range of 
$15 - 30$km s$^{-1}$ kpc$^{-1}$.
 In this work we assume $\Omega_p =26$km s$^{-1}$ kpc$^{-1}$ for the calculations.
Since the computational box is set in the reference frame of the 
spiral pattern the potential may be set as constant in time. 
The gravitational potential is also split into two components, 
the disk and the spiral arm, as $\Phi = \Phi_0 + \Phi_1$, respectively. We here 
 assume that the properties of the disk vary little at the 
scales of interest, therefore we neglect the gradient of $\Phi_0$ in the calculations 
as follows. 

The spiral pattern has been described in many previous works 
by a sinusoidal profile superimposed to the disk potential. 
A more consistent surface density distribution of the galactic disk though has been 
recently obtained for the Galaxy \citep{lep08}, based on stellar 
orbital velocities, showing that an exponential 
profile naturally arises from the linear theory of orbital perturbations. 
The excess in the surface 
density distribution was well described by a logarithmic spiral, 
with a Gaussian profile on azimuthal direction \citep[Eq.6 of][]{jun13}, as:

\begin{equation}
\Phi_1(R,\theta,z)=\xi_0 R e^{-\frac{R^2}{\sigma^2}\left[1-\cos(m\theta-f_m(R))\right]- \epsilon R -\left|kz\right|},
\label{eq:potencial}
\end{equation}

\noindent
where $\xi_0$ is the perturbation amplitude, $\epsilon$
is the inverse of the lengthscale of the spiral pattern, $\sigma$ the 
width of the Gaussian profile in the galactocentric 
azimuthal direction, $k = m/R\tan(i)$ the wavenumber, 
$i$ the pitch angle, and $f_m(R)$ the shape function, 
given by:

\begin{equation}
f_m(R)=\frac{m}{\tan(i)}\ln(R/R_i),
\end{equation}

\noindent
being $m$ the number of arms and $R_i$ the position 
where the arm starts.

For the Galaxy, the perturbation amplitude $\xi_0$ and the local 
surface density contrast between the spiral pattern and 
the disk ($\Sigma_{\rm a,max}/\Sigma_{\rm d}$), are related as:

\begin{equation}
\xi_0 \simeq 3.3\times 10^{3}\frac{\Sigma_{\rm a,max}}{\Sigma_{\rm d}} \frac{\sigma_\perp^2 \left|\tan{i}\right|}{m (R \sin{i})^2},
\end{equation}

\noindent
in km$^2$s$^{-2}$kpc$^{-1}$, for which the lengthscales of the spiral and the disk have been assumed 
as equals, and the width of the
Gaussian profile in the direction perpendicular to the arm $\sigma_\perp \equiv \sigma \sin i$ 
is defined.

%%%%%%%%%%%%%%%%%%%%%%%%%%%%%%%%%%%%%%%%%%%%%%%%%%%%%%%%%%%%%%%%%%%%%%%%%%%%%%%%%%%%%%%%
%		              Table - parameters                                        %
%%%%%%%%%%%%%%%%%%%%%%%%%%%%%%%%%%%%%%%%%%%%%%%%%%%%%%%%%%%%%%%%%%%%%%%%%%%%%%%%%%%%%%%%
\begin{table*}
\centering
\caption{Parameters used in each simulation. We explore the dependence with
the cloud-arm relative velocity, and properties of the arm potential, namely the width of the 
Gaussian profile $\sigma_\perp$ and its depth, related to $\xi_0$ (see Eq.\ref{eq:potencial}).
}
\begin{tabular}{ccccc}
\hline
Model &	$v_{\rm cloud,0}$(km/s) &	$\sigma_\perp$(kpc) &	$\xi_0$(km$^2$ s$^{-2}$ kpc$^{-1}$) & resolution (cells) \\
\hline
\hline
1  &	18.6  &	0.8  & $600$ & $1024 \times 1024 \times 512$ \\
2  &	18.6  &	0.4  & $600$ & $512 \times 512 \times 256$ \\
3  &	18.6  &	1.6  & $600$ & $512 \times 512 \times 256$ \\
4  &	18.6  &	0.8  & $3000$ & $512 \times 512 \times 256$ \\
5  &	18.6  &	0.8  & $600$ & $512 \times 512 \times 256$ \\
6  &	90.0  &	0.8  & $3000$ & $512 \times 512 \times 256$ \\
7  &	18.6  &	0.8  & $120$ & $512 \times 512 \times 256$ \\
8  &	4.6 &	0.8  & $600$ & $512 \times 512 \times 256$ \\
\hline
\hline
\end{tabular}
\label{tab:sims}
\end{table*}
%%%%%%%%%%%%%%%%%%%%%%%%%%%%%%%%%%%%%%%%%%%%%%%%%%%%%%%%%

Observationally, the arm-to-disk density ratio is determined as
$\Sigma_{\rm a,max}/\Sigma_{\rm d} = 0.13 - 0.23$ 
for spiral galaxies \citep{ant11}, which corresponds to 
$\xi_0 = 0.86 - 1.52 \times 10^3 (\sigma_\perp/R)^2$km$^2$s$^{-2}$kpc$^{-1}$, 
for $i=15^\circ$ and $m=2$. In our simulations, $\xi_0$ is 
kept as free parameter since we aim to understand the role 
of the spiral arm potential in the internal dynamics of 
the incoming gas. The sets of parameters used in our numerical simulations are given in Table \ref{tab:sims}. 
Most models were performed with a numerical resultion of $512 \times 512 \times 256$ cells, corresponding to a spatial resolution, in real units, of $\sim 1.95$pc. Model 1, which is mostly discussed along this manuscript was computed with $1024 \times 1024 \times 512$ cells, corresponding to $\sim 0.975$pc/cell.

The ambient gas is set, in the reference frame of the box, with initial velocity ${\bf v}_{bd}=\left(\Omega_0-\Omega_p\right)[R_0 - (y \cos i - x\sin i)]({\bf \hat{x}}\cos i +{\bf \hat{y}} \sin i )$, and gas density $n_0=1$cm$^{-3}$ at mid-plane ($z=0$), which 
exponentially decreases along vertical direction with scale height of 120pc. The temperature is set iinitially as uniform, being $T=10^3$K. 
Both Z-boundaries and upper Y-boundary are set as {\it open}. X-boundaries 
are set as {\it periodic}. Bottom Y-boundary 
is set as a constant inflow of insterstellar gas with constant density and velocity, given by the 
values previously described for the ambient gas.  
 
Finally, the ISM inhomogeneity that is interacting with the pattern potential is set as an overdense spherical cloud of radius $R_c=50$pc (as depicted in the right panel of Fig.\ref{fig1}), with uniform density $n_c=10^2$cm$^{-3}$ and temperature $T=100$K, and initially flowing the diffuse gas with a velocity 
amplitude given in Table \ref{tab:sims}. The cloud is initially positioned at the lower X-boundary. The dynamical evolution of such a cloud as it interacts with the arm is shown in the next section.

\section{Results}

\subsection{Simulations}

Let us start this section focusing on the results obtained for Model 1. We address each aspect of this specific model in detail here, which are then extended to the other models further in the paper to avoid 
unnecessary repetition. 

\begin{figure*}
{\centering
 \includegraphics[width=14cm]{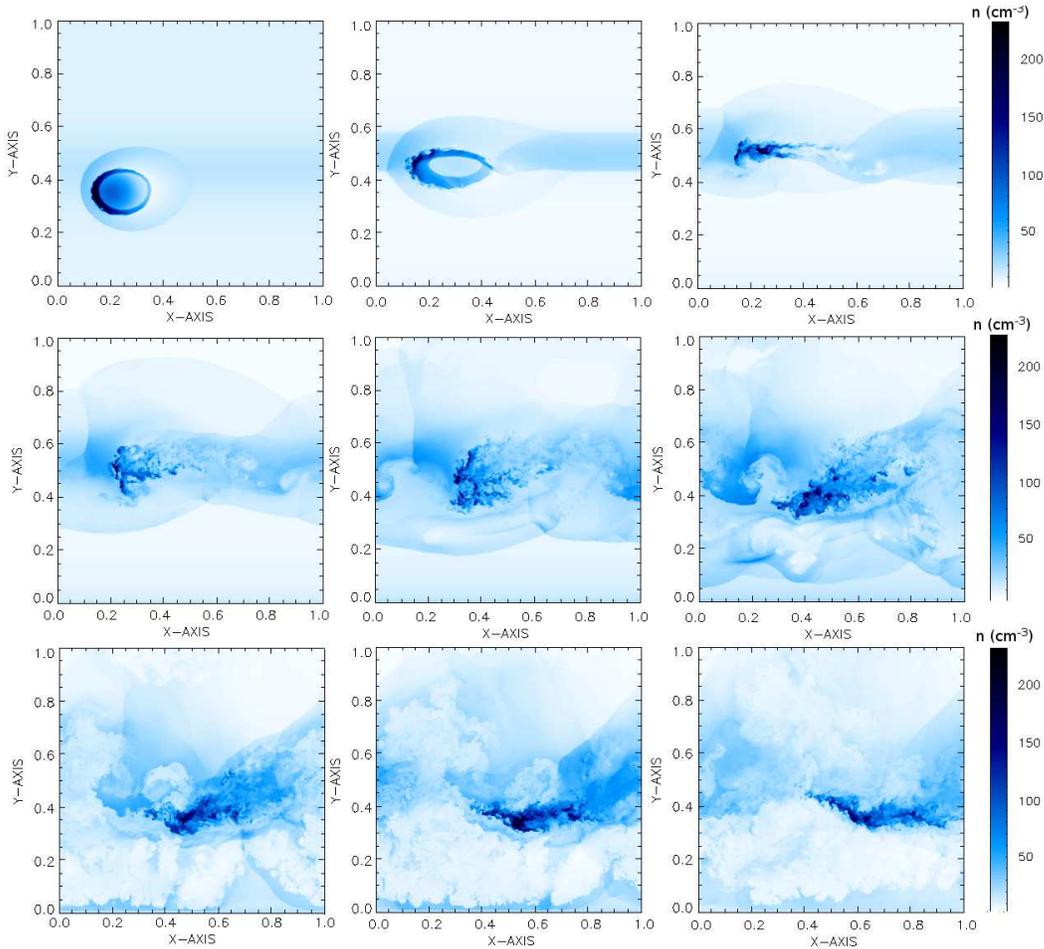}
}
 \caption{Mid-plane slices of density distribution of Model 1, at different times of the run. From top-left ($t=10$Myrs) to bottom-right ($t=90$Myrs), the time difference between snapshots is of $10$Myrs. Spatial axis in units of kpc.}
\label{fig:densitymaps}
\end{figure*}

As the cloud flows with the diffuse medium towards the spiral pattern it also interacts with the ambient medium. The relative velocity of the ambient gas and the cloud creates a weak shock that is clearly visible in the top-left image of Fig.\ref{fig:densitymaps}. The density is increased locally by a factor of $\sim 2.5$, and local shear instability is visible. Acoustic waves also propagate outwards, at the local speed of sound. 

The cloud does not fragment, or collapse, due to this effect though. On the contrary, at this stage ($t\sim 10$Myrs) the effects of the tides generated by the arm are already visible. Whilst shocked at the left border, by the incoming ISM gas from the lower-left boundary, the cloud is slightly stretched in the horizontal direction. The accelerated portion of cloud (upper part) then shocks with the gas within the arms potential at $t=20$Myrs (top-center image). The bottom part of the cloud, which is delayed, moves up more easily without shocking with any preexisting gas. The upper portion of the cloud is now slower and had lost considerable linear momentum to the ambient gas, while the bottom part of the cloud moves towards it. At $t=30$Myrs (top-right panel), approximately, the cloud material shocks with itself (collapse) resulting in a disk-like structure. At this stage parts of this system move in different directions, some upwards 
still sustaining the initial pull while other move downwards, falling back to the 
arm potential well after crossing it previously. Denser parts interpenetrate in several directions, resulting in the complex morphology seen in the mid-left panel of Fig.\ref{fig:densitymaps} ($t=40$Myrs). Now the homogeneous picture of the preset cloud is completely changed to a complex and turbulent-like morphology. In the following snapshots the cloud is seen fragmented, with the presence of clumps in a myriad of lenghtscales, being some of them denser than the original cloud. As follows, until $t=90$Myrs, the dynamics of the remainings of the cloud is dominated by the arm potential, with a bulk motion following the X-axis, i.e. along the orientation of the arm. The dense structures naturally diffuse as they interact with the ambient gas, and a stream of gas flowing to the top-right direction is visible. It is interesting to point out that in the timescales described here the gravitational collapse of some of the clumps could occur, however the collapse is not obtained here because self-gravity is 
neglected. A fraction of the gas mass would then be in stars before moving to the interarm region. The timescales during which the cloud remains trapped in the arm potential is discussed further in the paper.

\begin{figure*}
{\centering
 \includegraphics[width=14cm]{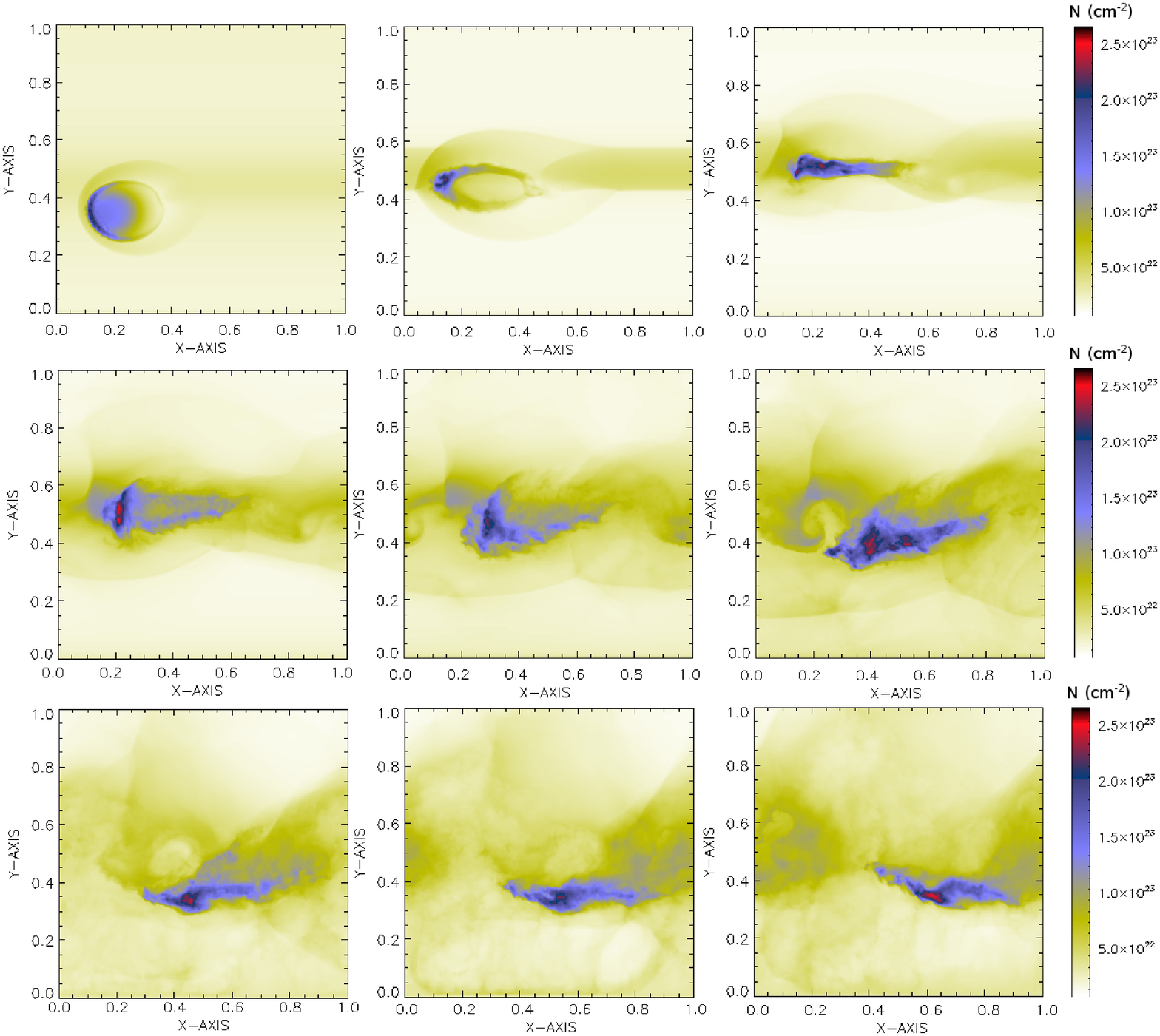}
} 
 \caption{Column density ($N$) distribution of Model 1, along the direction perpendicular to the galactic plane, at different times of the run. From top-left ($t=10$Myrs) to bottom-right ($t=90$Myrs), the time difference between snapshots is of $10$Myrs. Spatial axis in units of kpc.}
\label{fig:cd}
\end{figure*}

The column density maps for the line-of-sight (LOS) perpendicular to the disk plane are shown in Fig.\ref{fig:cd}. 
As described above based on the mid-plane density slices, the cloud is clearly stretched and collapsed into a planar structure up to $t=30-40$Myrs. In the column density projection it is possible to recognize that the sheared flows develop the KH instabilities, as clearly visible at $t=40$Myrs  and $50$Myrs (center row). The instability is possibly the cause of the fragmentation of the cloud. The final clumpy and filametary morphology is visible in the bottom row. Rayleight-Taylor instabilities are also easily recognized from the column density maps, as the ``voids", filled with hotter material, move both upwards and downwards (depending on its position with respect to the axis of the arm) away of the potential well of the arm. The cooled and denser material then follows the oposite trend, resulting in a number of filamentary structures perpendicular to the arm (and to the main body remanescent of the cloud). 

As the turbulence is supersonic at large lengthscales ($>10$pc), in the absence of any feedback from star formation, the cloud lifetimes is limited by the turbulent diffusion.
The turbulent eddy diffusivity is described as: 

\begin{equation}
D_{jj}=\frac{1}{2}\frac{\partial \Upsilon^2}{\partial t} \simeq |\delta v_l|l,
\end{equation}

\noindent
where $\Upsilon$ represents the averaged separation between pairs of fluid elements. 
For the turbulent properties of the cloud at the large scale $L\sim 100$pc, we found 
$D_{jj}\sim 10^{25}$cm$^2$s$^{-1}$, which result in a diffusion timescale 
$\tau_{\rm dif} \simeq L^2/D_{jj} > 100$Myrs. Therefore the diffusion timescale is longer than the time needed for the turbulence to be 
triggered as the cloud interacts with the arm. The total mass in the cloud decreases with time, reaching 
a value of $\sim 2\times10^5$M$_{\odot}$ at $t=100$Myrs. This is in agreement with the values obtained from the simulations, which could indicate that turbulent diffusion may be the dominant mechanism of disruption of the cloud. It is interesting to point here that the role of the inflowing diffuse gas on the diffusion/evaporation of the dense cloud is negligible.

\begin{figure*}
{\centering
 \includegraphics[width=12cm]{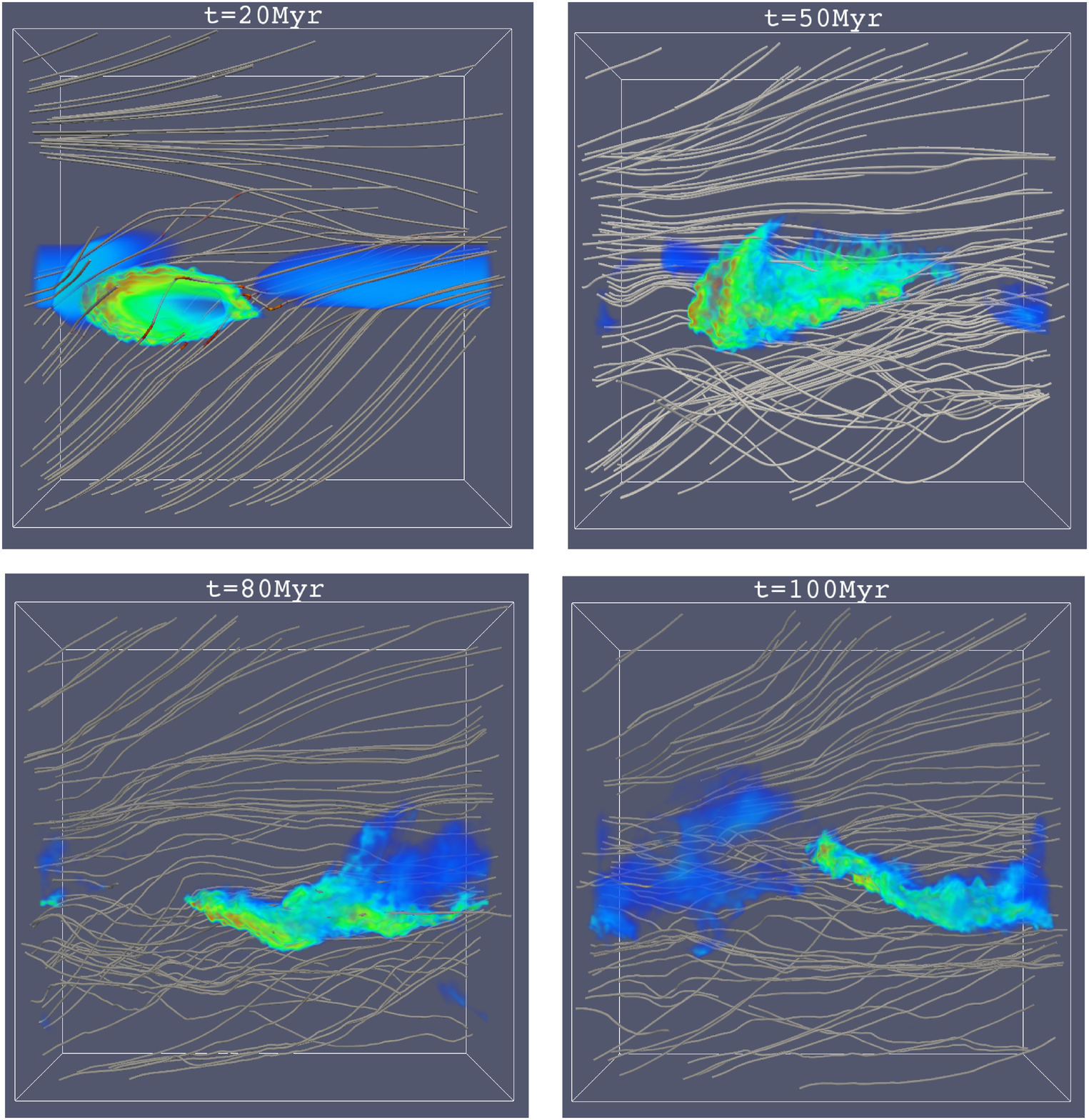}\\
 \includegraphics[width=12cm]{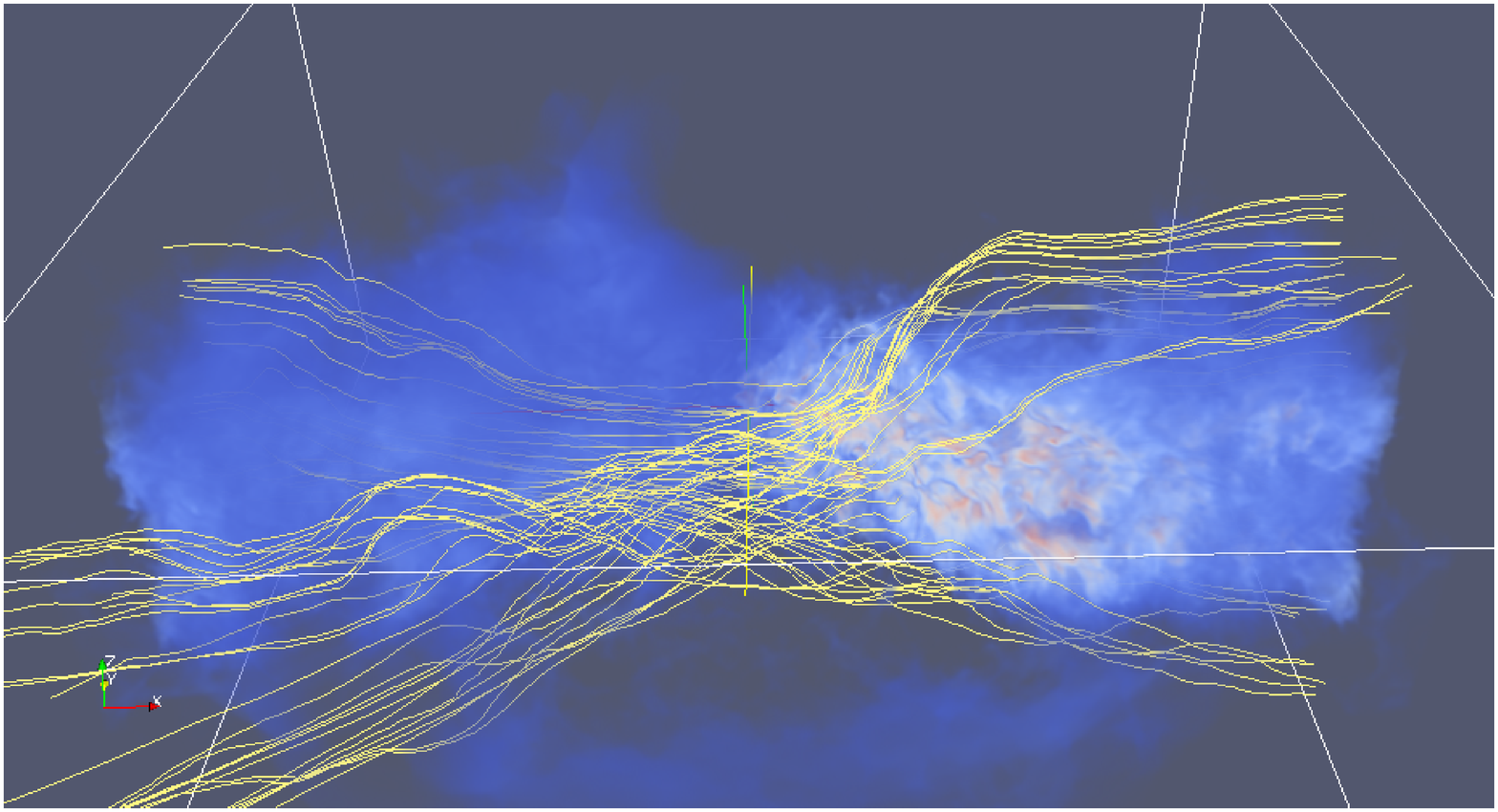}
} 
 \caption{3-dimensional projection of the density (colors) and velocity streamlines, for Model 1, at $t=20$, 50, 80 and 100Myrs. The streamlines represent the flow of the background diffuse medium. The denser cloud seems shielded to the flow. As the diffuse gas interacts with the cloud it is deflected and flow around the cloud. The diffusion of the clouds material is small revealing a low efficiency in driving motions within the dense structure. Bottom:  same projection but for a LOS inclined $30^\circ$ with respect to the galactic plane. Streamlines pass by, and not through, the dense structure.}
\label{fig:3d}
\end{figure*}

The inflowing gas can also be piled-up as it shocks with the cloud. Are the turbulent dense structures seen the result of original material perturbed by the arm, or turbulent gas resultant of shocked compression?  Unfortunately, in a grid-based code, as used in this work, it is not possible to flag particles and follow their positions during the run (e.g. as in SPH codes). We contour this issue by defining streamlines of fluxes instead. The 3-dimensional perspective of the cloud in combination with the velocity field, as streamlines, is shown in Fig.\ref{fig:3d} at four different times of the run.  

The streamlines have been selected to illustrate the the flow of the diffuse gas. Streamlines represent the orientation of the fluxes of matter connected through cells. In a standard head-on shock streamlines are perpendicular to the shock surface, showing that the incoming (upstream) flux results, after the shock, in a denser shocked (downstream) flow. However, in our simulations the streamlines are bended at the surface of the cloud.  Since the interaction is not a one-dimensional shock the oblique angle between the cloud surface and the velocity of the diffuse gas results in the deflection of the flow. The diffuse gas flows around the dense structure. Little, or virtually zero, diffusion of the low density gas into the cloud is observed. The dense gas is dynamically shielded from the shock with the ambient flowing gas. 

Even though a full Lagrangian integration of test particles would be required for a complete understanding of this process, the streamlines suggest that large scale, coherent and supersonic flows of diffuse matter are not dynamically relevant for the internal dynamics of the cloud. The main source of turbulence in our simulations is probably not due to the shocks between the cloud trapped in the spiral arm and the diffuse incoming flow. The internal dynamics of the dense gas must then be assessed.

\subsection{Statistics of turbulence}

\begin{figure}
{\centering
 \includegraphics[width=9.0cm]{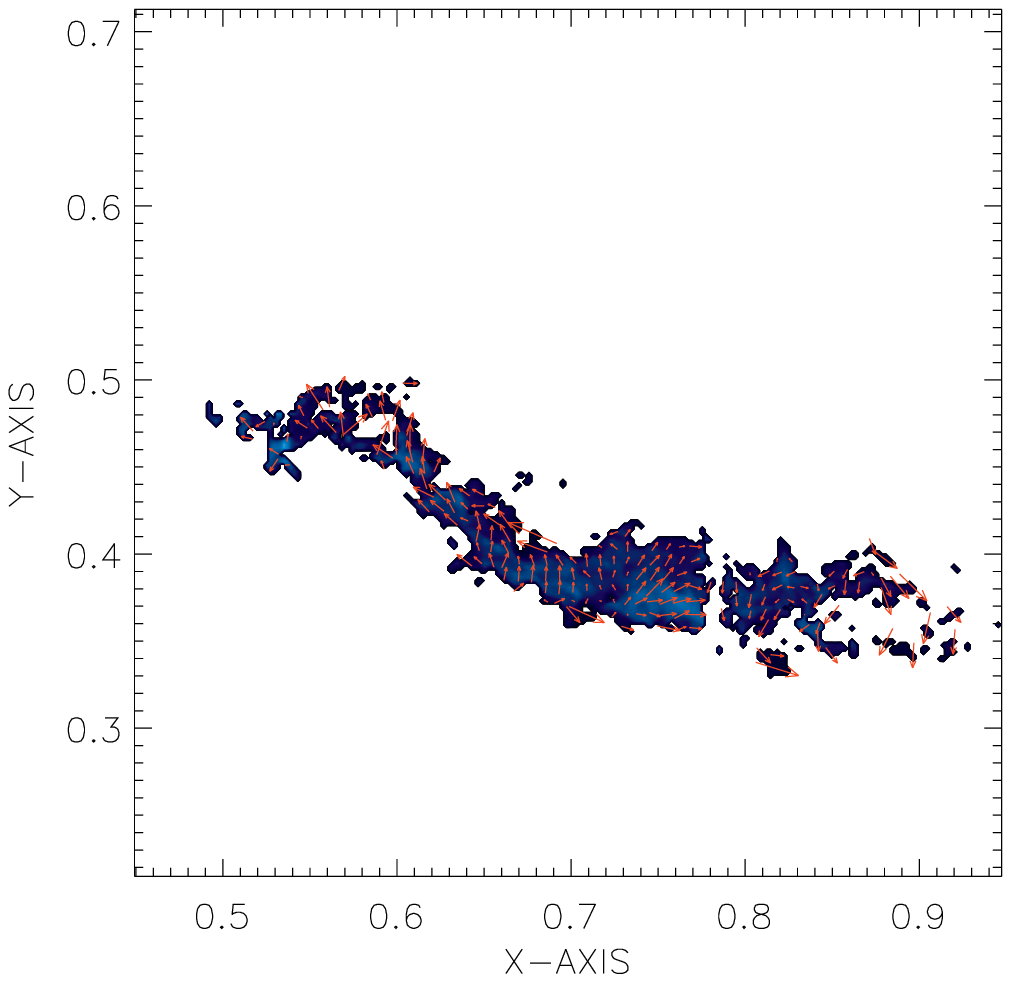}
}
 \caption{Internal velocity distribution of the dense structure, in the plane of the Galaxy,  for Model 1 at $t=100$Myrs. The field is obtained after subtraction of bulk velocity of the flow (see text), and turbulent 
 motions are highlighted.}
\label{fig:velturb}
\end{figure}

The internal motions of general flows may be complex, but not turbulent. Coherent flows in an ensemble of collapsing cores for intance may eventually look as turbulence from an observational perpective but naturally these are not turbulent in nature. Unfortunately, since local properties
of a turbulent fluid are unpredictable, turbulence can only be modelled
in terms of statistical quantities, mainly the velocity power spectrum or its correlation functions. 

\begin{figure*}
{\centering
 \includegraphics[width=7.0cm]{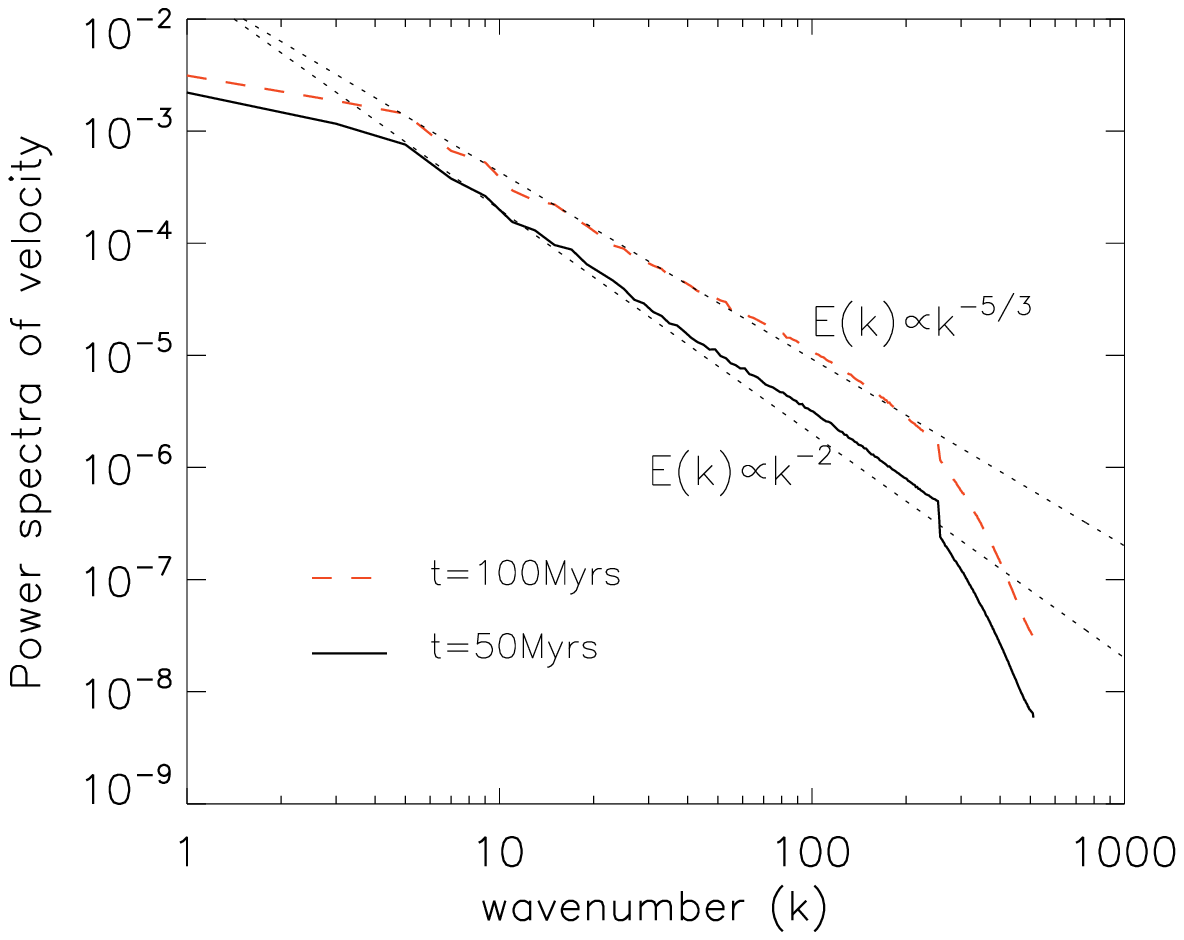}
 \includegraphics[width=7.0cm]{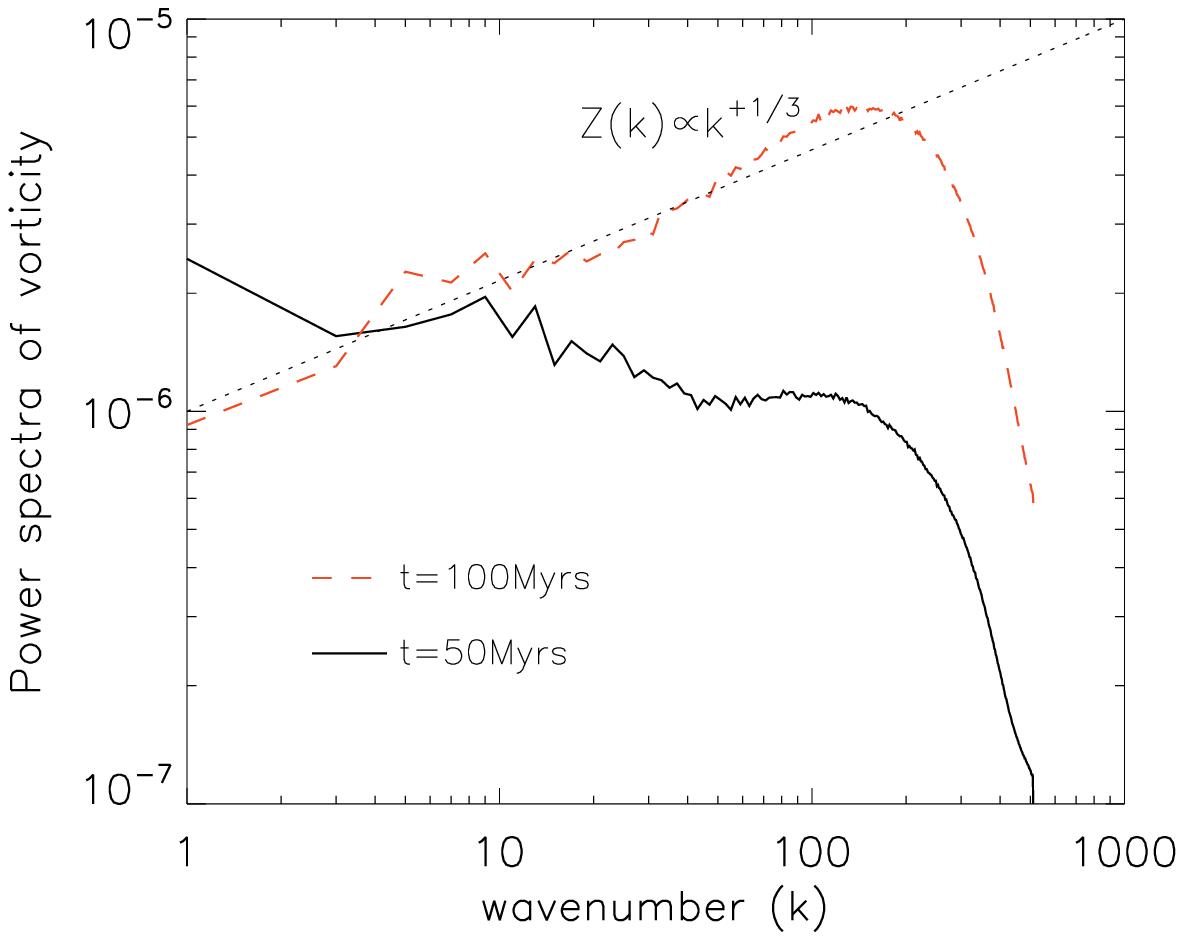}
}
 \caption{Power spectra of velocity (left) at two different times of the simulation, for Model 1, as a function of wavenumber $k=1/l$kpc$^{-1}$. The reference dotted lines corresponds to a standard Kolmogorov slope $-5/3$ and Burgers slope $-2$, which also corresponds to compressible turbulence. At first glance one may believe in a transition from compressible to incompressible 
 turbulence as time evolves. However, this is not the case as seen from the spectra of vorticity (right). The small power of vorticity at small scales reveal that most driving is then occuring at large scale. At $t=100$Myrs, the cascade is fully developed - as the typical timescale for the energy transfer though scales is $\tau \sim l/u_l \sim 3 \times 10^7 \mathcal{M}_L$yrs, at the largest scale -, and an increasing power spectrum of vorticity (with slope $\sim 1/3$, as in Kolmogorov case) is then observed.}
\label{fig:spectra2}
\end{figure*}

The velocity energy spectrum $E_u(k)$ is defined as $\int_{k=1/l}^\infty
E_u(k')dk'= \delta u_l^2$. If the turbulence is incompressible, isotropic and local\footnote{The concept of locallity here is understood in terms of the scales of the interacting waves. Triadic interactions of wave-wave interactions are of the type $\bf{k}_1+\bf{k}_2=\bf{k}_3$ (after selection rules have been applied to the Fourier transformed NS equation). If {\it local}, the energy transfer due to these interactions should peak at $k_1=k_2$}, the energy transfer rate between 
scales ($\epsilon$) may be assumed as constant. These conditions, proposed by Kolmogorov, naturally lead 
to the following spectrum for the velocity field:

\begin{equation}
E_u(k) \propto \epsilon^{2/3} k^{-5/3},
\end{equation}

\noindent
where, $\epsilon \simeq \delta
u_l^2/\tau_l$, where $\delta u_l$ is the velocity fluctuation
amplitude at lengthscale $l$, and $\tau_l = \tau_{\rm eddy} =
l/\delta u_l$ its dynamical timescale\footnote{Note that we 
  distinguish $\tau_l$ and $\tau_{\rm eddy}$ here, since $\tau_l$
  represents the timescale for energy transfer at scale $l$, while
  $\tau_{\rm eddy}$ is the eddy turnover timescale. In the theory of Kolmogorov
  both timescales are equal, which is not true in general,
  e.g. in magnetized plasmas \citep[see][]{fal14}}. 

The energy spectrum of velocity was obtained in the simulations after 
removal of the bulk motion of the system with respect to the arm (see Fig.\ref{fig:velturb}). In order 
to reduce the contamination of the sheared velocity profile - in the reference 
frame chosen - at large scales, we subtract 
the average velocity obtained at each radius $R(x,y)$ from the velocity 
of the cells it intercepts. The remaining velocity distribution is a composition 
of coherent flows dynamically related to the local gravitational potential, 
the shocks, and the turbulence itself. The result is shown in Fig.\ref{fig:spectra2} 
(left) for $t=50$Myrs and 100Myrs.

The two spectra are slightly different with the earlier velocity field ($t=50$Myrs) 
showing a steeper spectrum, comparable to a power-law with slope $\sim -2$. Such a steep power spectrum 
is expected for compressible turbulence with large Mach numbers ($M_s = <\delta 
u^2>^{1/2} / c_s \gg 1$). At $t=100$Myrs, the spectrum reveals a more Kolmogorov-like 
slope. Such ``evolution" of the turbulent pattern of the flow may indicate that the 
origins of turbulence occur in a highly compressible flow, such as strong cooled 
shocks, and then evolve towards a less drastic situation as energy is dissipated. 
Such a case would be in agreement with that of colliding-flows in the ISM. A strong 
radiative shock, i.e. highly compressible, would drive supersonic turbulence at 
large scales. However, as pointed in the Introduction, such models are unable to 
provide either supersonic turbulence nor large scale turbulence. The typical driving 
for colliding flows, for instance, occurs at the shock width scales and with amplitudes $M_s \sim 1$.
Another explanation for the observed slope is that turbulence is being driven at 
large scale but have had not enough time to cascade towards small scales. The 
steeper spectrum would then be caused by a lack of energy at small scales, instead 
of a more efficient cascade process due to compressibility. The typical timescale 
for turbulence to cascade at a given lengthscale $l$ is $\tau \sim l/u_l \sim l^{2/3}L^{1/3} \left(M_{s,L} c_s\right)^{-1}$, being $M_{s,L}$ the sonic Mach number of the turbulent motions at the largest lengthscale $L$. For a temperature of $T=10^2 - 10^3$K, and a scale of $L \sim 100$pc, we find $\tau_L \sim 3 - 5 \times 10^7 M_{s,L}^{-1}$yrs. For a mildly supersonic turbulence the turbulent cascade should not be fully evolved at $t<50$Myrs, in agreement to the proposed idea.

Another tool that may be used to identify if the turbulence is actually driven at large scale is the statistics of enstrophy ($Z_u(k)$), defined as the energy spectrum of vorticity ${\bf \omega}={\bf \nabla} \times {\bf u}$, i.e. $Z_u(k) \equiv 2\pi k \left\langle \left|\tilde{\omega}({\bf k})\right|^2\right\rangle$. Similarly to the energy spectrum of velocity, enstrophy is related to the second momentum of vorticity distribution as $\int_{k=1/l}^\infty
Z_u(k')dk'= \omega_l^2$.  
The enstrophy is an invariant of the Navier-Stokes equation in 2-dimensional turbulence, with interesting implication for the galactic disk case if motions perpendicular to it get constrained somehow, specially for the inverse cascade problem (see Section 4). In any case, even in the three-dimensional case, 
the spectrum of vorticity can lead to interesting conclusions. In a Kolmogorov like turbulence, since $Z_k = k^2 E_k$, the slope of enstrophy is $+1/3$, resulting in the acumulation 
at small scales. The peak in $Z_k$ is expected to occur at the transition scale where dissipation starts to dominate the dynamics of the flow. If the driving source was shock induced at scales as 
large as the shock widths, and inverse cascade operate, the observed enstrophy 
spectra would be peaked at small scales - during the whole simulation - and slowly evolving in time towards the large scales. This is exactly the opposite to 
what is shown in Fig.\ref{fig:spectra2} (right). There the enstrophy spectrum at 
$t=50$Myr is not peaked at small scales, but is flat. At $t=100$Myr most of the 
curve is nicely reproduced by a $+1/3$ power-law, at the inertial range, with a peak 
at small scales. The total enstrophy spectrum increases with time, specially at small scales. The increasing pile-up of power at the small scales indicate a ``cascade" process, which is associated to the fact that the driving of the turbulence must have occured at larger scales than that of shocks.

\begin{figure*}
{\centering
 \includegraphics[width=14cm]{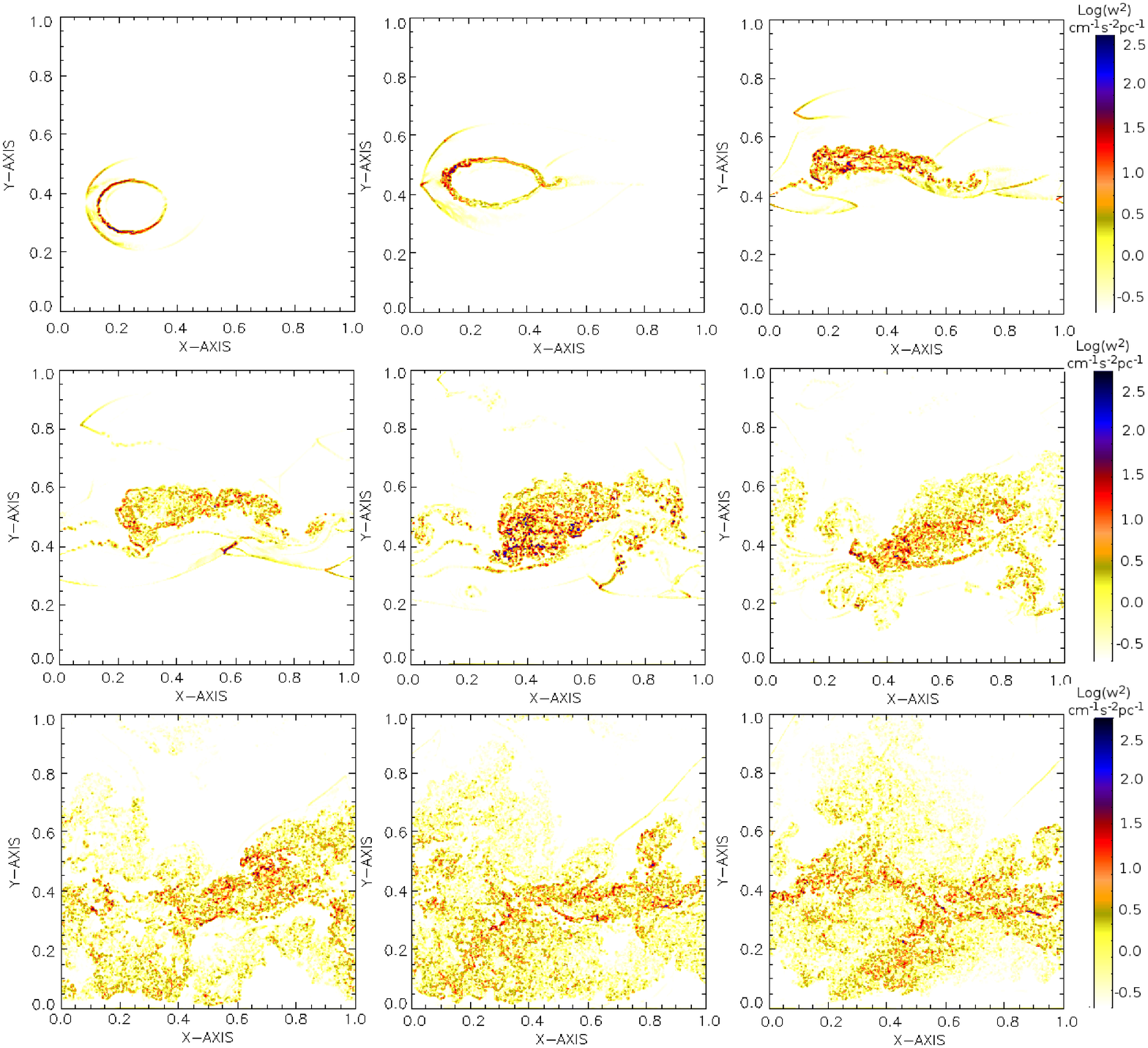}
} 
\caption{Mid-plane maps of the squared specific vorticity ($w^2 \equiv n\omega^2$) obtained for Model 1, at different times of the run, from top-left ($t=10$Myrs) to bottom-right ($t=90$Myrs). Spatial axis in units of kpc. Three different moments are identified from these plots. First at $t= 10 -20$Myrs, in which turbulence is driven at the border of the cloud at the two shock surfaces due to local instabilities. The second is when the whole cloud becomes turbulent, at $t\sim 30-40$Myrs. The third is when most of the computational domain shows 
strong vorticity at $t>80$Myrs. }
\label{fig:ensmaps}
\end{figure*}

Figure \ref{fig:ensmaps} shows the spatial distribution of the squared vorticity in the galactic plane, at different times of the run, from top-left ($t=10$Myrs) to bottom-right ($t=90$Myrs). From this it is not clear though {\it on what scales} turbulence is driven, as its power peaks at the dissipation scales. On the other hand it is possible to use this quantity as an indication of {\it where} turbulent motions have been excited. 
The early stages of the cloud-arm interaction, at $t= 10 -20$Myrs, are dominated by shock-induced vorticity. These are caused by both the shear of the downstream and upstream flows, as well as to the local instabilities, as discussed before. These regions are restricted to the shocks though and had not enough time to diffuse, or to cascade inversely, in the dynamical time of the cloud. It can be seen that at  $t= 30 -40$Myrs the whole cloud has become filled with increased vorticity. Notice that the triggering mechanisms here are different. At the later stage the cloud has already been tidally stretched and contracted, as it is now falling back to the arm after one crossing. At $t>60$Myrs the surrounding diffuse medium also presents increased vorticity. We find the main driving mechanisms to be the nonlinear interactions of waves excited by the shaky cloud, during its motion in and out of the spiral arm, and the Rayleigh-Taylor instability, as it advects part of the turbulent gas out of the arm.

\subsubsection{Probability distribution of velocity lags: intermittency}

Turbulence is understood as self-similar for all scales. This is obviously not completely true 
since self-similarity must break as we get close to the dissipation scales. Similar 
behavior is expected if the statistics of coherent and long-lived structures may not 
be neglected. This is the case, for instance, of supersonic turbulence where shocks 
generate structures that may be (and generally become) decoupled to the surrounding 
ambient turbulence \citep[see e.g.][]{fal11}. As intermittency is characterized by 
coherent structures that break the self-similarity of turbulent chaotic motions it may be 
detected in turbulent flows as a departure from Gaussian distributions at small scales.

\begin{figure}
{\centering
 \includegraphics[width=7.0cm]{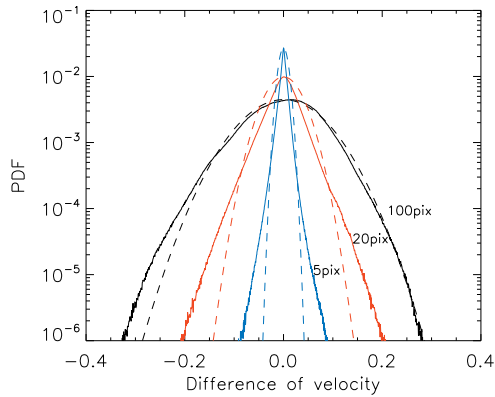}
}
 \caption{Probability distribution functions (PDFs) of $\Delta v_l 
 \equiv {\bf v(r)}-{\bf v(r+l)}$, obtained from Model 1 at $t=90$Myrs, for three different 
separation lengths, $l=5, 20$ and $100$pixels, which correspond to $\sim 4.9$pc (blue), $19.5$pc (red) and 
$97.6$pc (black). Dashed lines correspond to Gaussian fits. At large scale (black line) the lag of velocity shows an broad wing at negative velocities. This is caused by the large scale galactic flow. A symmetric departure from Gaussian distributions is more obvious at small scales though, characterizing the intermittent behavior of the turbulence in our model.}
\label{fig:lag}
\end{figure}

We computed the probability distribution functions (PDFs) for the variable 
$\Delta v_l = {\bf v(r)}-{\bf v(r+l)}$, known as the ``velocity lag" 
for the lenghtscale $l$. The results for Model 1, at $t=90$Myrs, are shown 
in Fig.\ref{fig:lag} for three different separation lengths, $l=5, 20$ and 
$100$pixels, which correspond to $\sim 4.9$pc (blue), $19.5$pc (red) and 
$97.6$pc (black). Each of the PDFs was fitted by a Gaussian distribution, 
overplotted as dashed lines. None of the distribuions showed significant skewness 
(3rd moment), which is reasonable for an isotropic distribution. An exception naturally 
arise due to the large scale galactic flow profile which results in the excess at negative 
velocity lag (black line). The PDFs of the smaller scales are symmetric though. 

The kurtosis (4th moment), 
on the other hand departures from Gaussian with increased kurtosis for smaller lags ($l$). 
At smaller scales the distributions are more peaked with extended tails on both sides. 
The intermittent behavior here may be understood as being caused by shocks and/or sheared motions. 
As dense structures may be shielded to diffuse gas inflowing in its direction this results in an excess of material at larger velocity shifts at the expense of the peak of the distribution. The few and 
coarsely distributed small clumps are not able to modify the statistics at larger scales though, which 
show a quasi-Gaussian distributions. Shocks would present similar properties, as the velocity field 
suffers sharp fluctuations in amplitude within scales as small as the shock width. It is difficult to determine 
the dominant process in our models since both processes occur, however we may point out that the shocks 
observed in the simulations are much narrower than the scales were the intermittency is observed. 
The clumps, on the other hand, are predominantly at similar scales of tens of parsec, and are
 more likely to be causing the intermittent behavior.

The characterization of the intermittency is of particular interest here in order to understand the 
origins of turbulence in the ISM. The inverse cascade turbulence is known to present little, 
or virtually zero, intermittency \citep[see][and refs. therein]{bof12}. The strong intermittent behavior
in our model at small scales is another indicative that the inverse cascade cannot be the dominant driver of the observed turbulence.

\subsection{Larson's scaling relation}

From the observational point of view it is not possible to access the three-dimensional information 
of the interstellar turbulence. We must always be aware of projection effects, as the signal is 
integrated over considerable lenghtscales, much larger than the dissipation scales in most 
cases. Let us reconstruct observable quantities from the simulated cube here. 

We follow the 
approach of \citet{fal11} to calculate the synthetic observational velocity dispersions, and compare 
to those obtained from the actual three-dimensional distribution.
The three-dimensional velocity dispersion is a function of the scale $l$. For each $l$, the 
computational domain is divided in $N_l$ subvolumes $\mathcal{V}_{\rm 3D} = l^3$. The dispersion of 
velocity for the size $l$ is then obtained as the mean value of the local density-weighted velocity 
dispersions ($v^*=\rho v$) obtained for each subvolume. 
For the synthetic observational dispersion, on the other hand, 
we subdivide the plane representing the sky in squares of area $l^2$, which mimics the 
observational beamsize. Here we chose a LOS along x-direction, i.e. an observer within the 
galactic plane looking through along the spiral pattern.
The dispersion of $v^*$ is then calculated within each of the subvolumes 
$\mathcal{V}_{\rm proj} = l^2L$, as a function of $l/L$. 

\begin{figure}
{\centering
 \includegraphics[width=7.0cm]{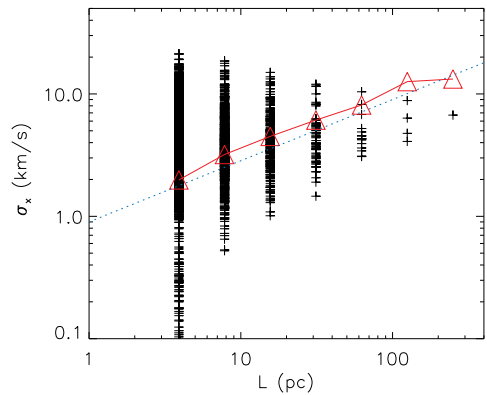}
}
 \caption{Average 3-dimensional density weighted velocity dispersion (red triangles), and the synthetic observational values (black crosses) for all LOS's defined for each lengthscale, or beamsize, as a function of the lengthscale. Large dispersion of synthetic values at small scales is expected in highly structured density 
 distributions, e.g. in supersonic turbulence. Dotted blue line is the Larson scaling relation $\sigma_v \sim 0.9 l^{0.5}$km s$^{-1}$.}
\label{fig:larson}
\end{figure}

The results of both calculations are shown in Fig.\ref{fig:larson}, where the averaged 3-dimensional density weighted velocity dispersion are shown as red triangles, while the synthetic observational counterparts are shown as black crosses, for all LOS's defined for each lengthscale, or beamsize, as a function of the lengthscale. At large scales both values converge, while at small scales a large dispersion of the synthetic observational values is observed aroud the expected 3D measure. This behavior occurs if the turbulent flow presents a highly structured density distribution, e.g. in supersonic turbulence. Voids and multiple overlaid 
dense structures at different LOS's, for the same beamsize, result in very different velocity dispersions. 

Historically, observational surveys of the ISM revealed a more linear scaling relation, which led to the 
direct fit of the Larson scaling relation \citep{lar81} $\sigma_v \sim \sigma_{v_0} (l/l_0)^{\alpha}$. Such 
fit is supported by the theory itself, to some extent, due to the relation between the velocity dispersion and the structure function of the turbulent distribution. From the theoretical point of view,
the energy spectrum ($E_k$) is equivalent - though in Fourier space - to the second order 
structure function. The structure function, or two-point correlation function is defined as:
  
\begin{equation}
S_p(l)=\left\langle \left\{\left[ \textbf{u}\left( \textbf{r} + \textbf{l} \right) - \textbf{u} \left( \textbf{r} \right) \right] \cdot \textbf{l}/l \right\} ^p \right\rangle \simeq C(p) \epsilon^{p/3}l^{p/3},
\end{equation} 

\noindent
where $p$ is a positive integer representing the moment order and ${\bf l}$ is
the vector lag in space. For a Kolmogorov turbulent spectrum, 
the second-order ($p=2$) SF is then $S_2(l) \propto l^{2/3}$, and 
therefore:

\begin{equation}
<\delta u_l^2>^{1/2} \simeq (\epsilon l)^\frac{1}{3}.
\label{eq2}
\end{equation}

Observations of several molecular clouds in the Galaxy indicate however a fiducial fitting for each of these as $\sigma_v \sim 0.9 l^{0.5}$km s$^{-1}$ \citep[see][]{hey04}. The observational slope differs slightly from what is expected for a Kolmogorov scaling. If the complete subset data from these several clouds are plotted together instead, i.e. not only averaged values for each scale, the linear relationship proposed by Larson is less evident and a picture similar to that shown in Fig.\ref{fig:larson} is observed \citep{bal11}. Possibly the difference between the slopes obtained for individual clouds and those expected for a turbulent flow may be related to observational issues, and not to the local nature of the gas.
What is particularly important here is that our models present turbulent amplitudes in agreement with those observed in molecular clouds. 

\subsection{Other models}

The results shown so far have been obtained based on Model 1, which has been chosen as 
the fiducial model, given its higher numerical resolution and its initital setup. In order 
to understand how, or if, different properties of ISM/spiral arm would change the conclusions 
made so far we will now analyse the dynamical evolution of the other models.

All models, except for Model 3 (see Table 1), are very similar in their general behavior. 
The cloud is initially moving towards the arm, which pulls the cloud closer accelerating and 
stretching it. As the cloud passes through the arm the internal motions become more complex. 
Internal shear and shocks dissipate part of the kinetic energy gained from the potential 
of the arm, triggering the turbulence within the cloud. The surrounding medium is also pushed as 
the cloud moves, triggering turbulence in the diffuse gas as well. The loss of kinetic energy 
makes the clouds, which were initially freely moving in the galactic disks, bound to the arm's 
potential well. 
Model 3 is the only model where the cloud is unbound after its crossing. Though 
random motions are still driven within the cloud, the amplitude of the perturbations observed 
 is smaller compared to the other models. The cloud leaves the box at $t\sim80$Myrs.

\begin{figure}
{\centering
 \includegraphics[width=9.0cm]{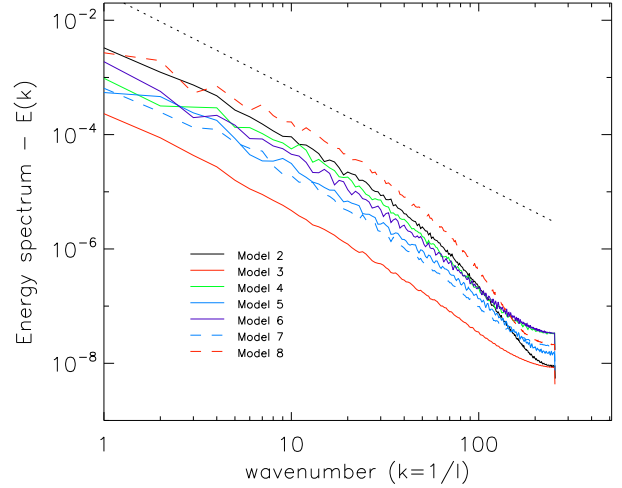}
}
 \caption{Energy spectra of Models 2-8, with a Kolmogorov -5/3 slope spectrum as reference line. All models 
 present similar spectral distributions, with small differences in amplitude (less than an order of magnitude), except for Model 3 where the amplitude is much smaller than the statistical average. This model presents a cloud that leaves the box after its interaction with the arm. The remaining turbulence is due to the disturbed ambient gas.}
\label{fig:spectra}
\end{figure}

In Fig.\ref{fig:spectra} we present the energy spectra obtained for Models 2-8, at the last 
snapshot. All spectra present similar profiles, though with shorter inertial range when compared 
to that of Model 1, due to the coarser grid used in these models. An inertial range of about one 
decade in wavenumber is observed in all models, with slopes similar to $-5/3$. Compared to each 
other the models present similar amplitudes, except to Model 3, which lies far 
 below the average. The outlier situation of Model 3 is explained since this is the only model run 
 where the cloud leaves the box through the upper Y-boundary after interacting with the spiral pattern. 
 The cloud is unbound to the gravitational potential of the arm. Notice that other examples are also 
 initially unbound, but the internal dissipation during the interaction removes enough linear and angular 
 momentum of the cloud that then becomes bound the the arm (at least long enough for a second interaction). 
 
\begin{figure}
{\centering
 \includegraphics[width=9.0cm]{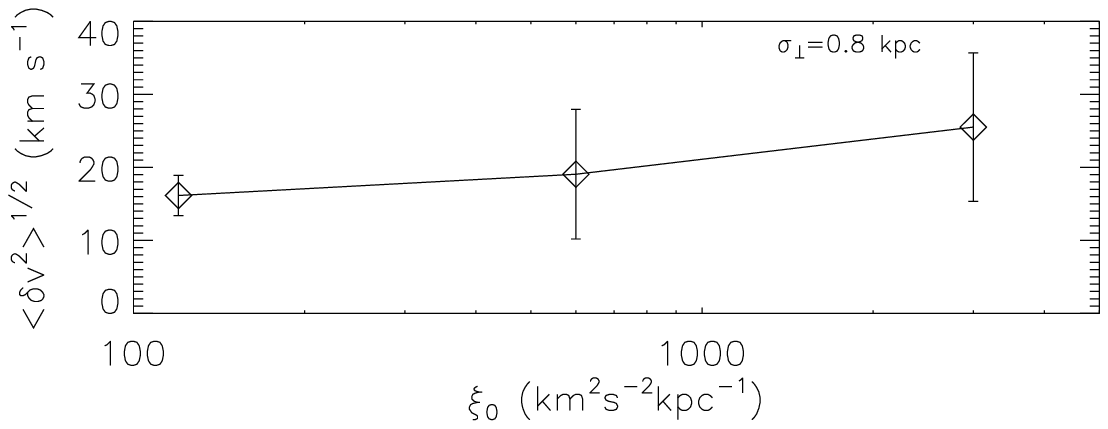}
 \includegraphics[width=9.0cm]{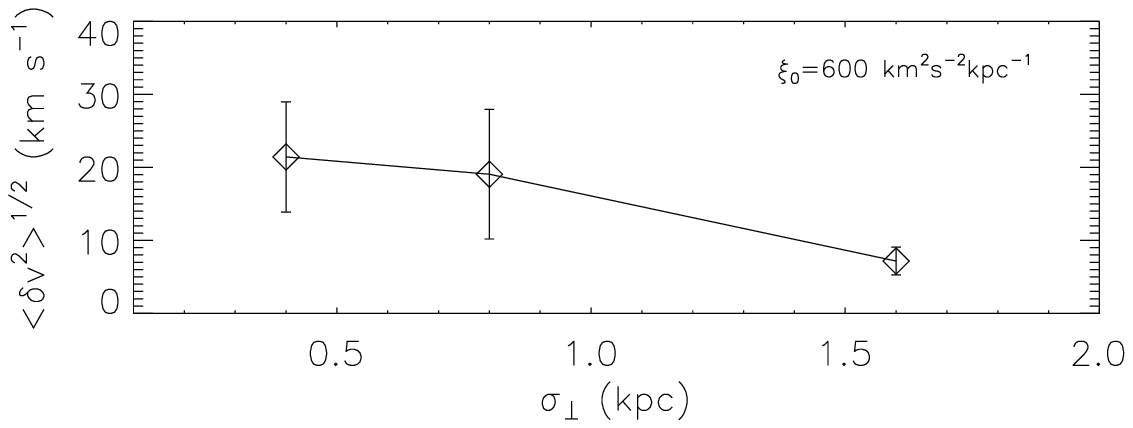}
 \includegraphics[width=9.0cm]{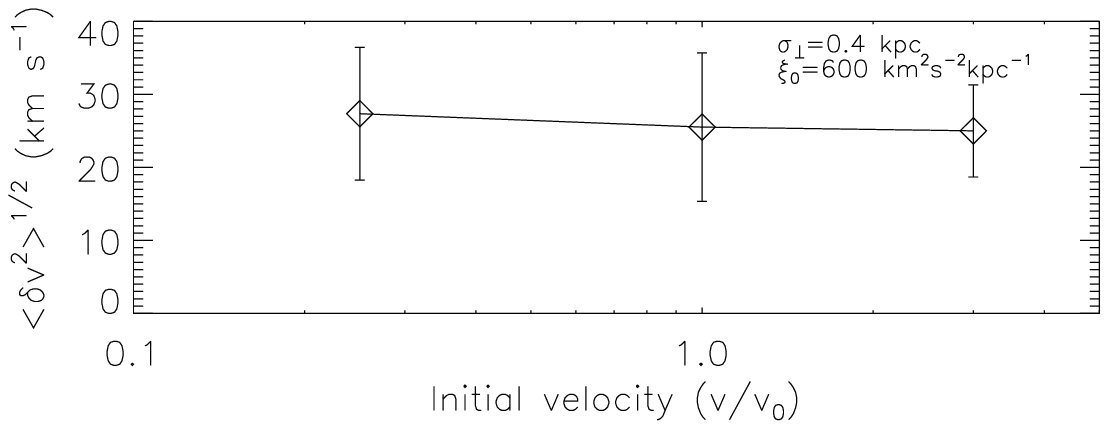}
}
 \caption{Velocity dispersion of turbulence, averaged over the last 50Myrs of the simulations, 
 as a function $\xi_0$ (top), $\sigma_\perp$ (middle) and relative velocity of the spiral pattern 
 with respect to the rotating gas (bottom). For the latest, the value of reference is $v_0 = (\Omega_0 - \Omega_p)R_0=40$km s$^{-1}$. No correlation is observed between the turbulence driven and the 
 initial cloud-arm relative velocity. Turbulence amplitude is related to $r_0/\sigma_\perp$ and $\xi_0$, which 
 is related to the surface mass density of the spiral pattern.}
\label{fig:corrs}
\end{figure}

For the models where the cloud becomes bound to the spiral arm the turbulent velocity dispersion may be related to other local properties. We computed the turbulence amplitude $<\delta v^2>^{1/2}$ for all models, at different times, which was then averaged over the different snapshots. The correlations of the averaged 
dispersions found with respect to the initial parameters $\xi_0$, $\sigma_\perp$ and initial cloud-arm relative velocity $v$, are shown in Fig.\ref{fig:corrs}.

The parameter $\xi_0$ is related to the surface mass overdensity of the spiral arm, while $\sigma_\perp$ 
its distribution. These two parameters reveal the strength of the gravitational forces acting in the 
cloud. On the other hand $v$ is not related to the gravitational forces, but is the parameter that controls 
the shock strength between the cloud and the arm.

It is interesting to notice from Fig.\ref{fig:corrs} that there is no clear correlation between the turbulence 
driven in our models with the initial relative velocity between the cloud and the arm. The strength of the shock should be, if the NTLI was the dominant process in 
driving random motions in the gas, strongly correlated to the level of turbulence. This is clearly not observed 
here. On the other hand the level of turbulence seems to be exclusively related to the properties of the arm instead, such as its surface mass density and its compactness. This surprising result is analyzed in details in the next section.

\section{Triggering mechanisms: instabilities, scales and turbulent amplitudes}

The models presented in this work were successful in reproducing many aspects of the turbulence in the interstellar medium, namely i) the universality, as the mechanism that operates here is generalized to the entire galactic disk, it ii) provides turbulent amplitudes at large scales ($>100$pc), iii) with amplitudes of $>10$km s$^{-1}$, and iv) it results in full spectra of velocity. It is not clear yet though what physical process is dominant in driving the turbulence in the models.
In this section we address and evaluate the processes that may be the main driver of the turbulence in our models.

\subsection{Instability-Driven Small Scale Turbulence}

Shocks induce structuring of the gas mostly due the non-linear thin-layer instability (NTLI), which also 
triggers Kelvin-Helmholz (KH) at similar scales, resulting in a well developed turbulent flow. 
Despite the apparent relevance of gas-arm shocks on driving galactic scale turbulence, 
the results obtained from the numerical simulations show a different scenario, where large scale effects dominate. 

The scales at which these instabilities take place are small compared to 
the lengthscales of the system. In the simulations presented in this work a cloud 100pc wide 
interacting with a spiral arm shows shock induced turbulence at its edge, i.e. the 
working surface, exclusively. The turbulence driving therefore occurs at small 
scales ($l < 10$pc), and not over the whole volume of the cloud. 
One would then expect that, unless the cloud is completely disrupted into 
small fragments of the same size of the driving scales, two processes must 
then occur - not specifically in this order - to generate a fully developed 
turbulent cloud: i) the diffusion of the turbulent energy through the 
whole volume of the cloud, and ii) an inverse cascade of the turbulence 
from small to large scales. As explained below, these two steps impose 
serious restrictions to the model of shock induced turbulence in the ISM.

Let us first focus on the filling factor of the energy injection. 
Waves excited by the nonlinear evolution of the shocked cloud (NLTI), with 
amplitudes $v_{\rm inj}$, could eventually propagate inwards and result in 
a fully turbulent cloud. 
The turbulent diffusion of transonic perturbations ($v_{\rm inj} \sim c_{\rm s}$) 
over the entire cloud ($V \sim L^3$) occurs on a timescale of $\tau_{dyn} 
\simeq L/c_{\rm s} \sim 20-80$Myr. However, because of the isotropic dillution 
of kinetic energy over the whole volume, in order to keep the turbulence amplitude 
as large as that driven initially, the cloud must interact with the arm for a 
time longer than $\tau_{\rm cross} > \tau_{dyn}/f$, where $f$ is the volume 
filling factor of the shocked region (neglecting any loss of kinetic energy). 
Even overestimating the shock thickness as $\Delta_{shock} \sim 0.1 L$, one obtains 
$\tau_{\rm cross} >200$Myr, which is probably too large compared to the dynamical 
timescales of cloud-arm interactions, or even compared to the lifetimes of these objects.

\subsection{Inverse Cascade: from small to large scales}

The second main issue regarding shock induced turbulence arises from the fact 
that the ISM turbulent spectrum peaks at scales of hundreds of parsecs. 
Shock induced turbulence is characterized by the transfer of energy and mometum 
from large scale (and coherent) converging flows into a multi-scale, chaotic and diffusive 
field. This phenomenon has been identified in numerical simulations \citep[e.g.][and others]{hunter84,walder2000,aud05,hei2005,vs2006,vs2007,hei11,inoue2013,folini2014}, being promptly 
related to the collapse of dense strucutures in the ISM and star formation, and 
analytically described by \citet{vish94} as the nonlinear evolution of perturbations 
in shock-bounded slabs. 

The NTLI is understood to arise at large Mach number shocks as perturbations perpendicular to the working surface of the shock may grow nonlinearly. The growth rate $\nu$ of surface bending perturbations ($\delta$) is $\nu \sim c_{\rm s} k (k\delta)^{1/2}$. The perturbations at smaller scales therefore grow faster, and drive local vorticity at scales as large as the shock thickness ($\delta \simeq \Delta_{\rm shock}$). Quenching should occur when the local turbulent kinetic pressure start acting as restoring force. The velocity dispersion at the slab is then expected to saturate around the local sound speed, i.e. the amplitude of driven eddies 
$v_l \simeq c_{\rm s}$. The statistics of the shock bound region of colliding flows has been recently studied by means of 3-D numerical simulations by \citep{folini2014}, which confirmed the low efficiency in the conversion of kinetic energy into turbulence. Similar results were obtained in the magnetized case from 3D MHD simulations by \citet{fal12}. 

%The Navier-Stokes equation for a viscous fluid, with explicit friction and external forcing terms, is given as:

%\begin{equation}
%\frac{\partial {\bf u}}{\partial t}+{\bf u} \cdot {\bf \nabla u} = \frac{\nabla p}{\rho} + \nu \nabla^2 {\bf u} - {\bf F}_{\rm friction} + {\bf F}_{\rm forcing},
%\end{equation}

%\noindent
%where $\nu$ and $\alpha$ represent the viscous and friction coefficients, and ${\bf F}_{\rm forcing}$ a forcing %term. We also define the vorticity as $\omega \equiv {\bf \nabla} \times {\bf u}$. 

\citet{kra67} noticed that the dissipationless/unforced 2-dimensional Navier-Stokes equation admits the energy ($E_k$) and the enstrophy ($Z_k$) as quadratic invariants. Specific cascades for each of these is therefore expected. Once driven, at an intermediate scale $l_{\rm inj}$, it is possible to show that enstrophy cascades to small scales while energy should present an inverse cascade \citep[see review by][]{bof12}. At equilibrium, the energy is dissipated at the smallest scale ($l_\nu$) due to viscosity, and at the largest scale ($l_\alpha$) due to the dynamical friction of eddies. The slopes of power spectra are derived as $-5/3$ for the inverse cascade, and $-3$ for the direct cascade. Such inverse cascade has been observed in laboratory experiments, presenting scalings similar to those of \citet{kra67} \citep[e.g.][]{paret98}.
This picture may be different in 3-dimensions though as enstrophy is no longer invariant (due to the non-linear term $\left({\bf \omega} \cdot {\bf \nabla}\right) {\bf u}$ of the Navier-Stokes equation, known for the process of vortex stretching). One might especulate if both inverse and direct cascades should occur simultaneously in 3-dimensional turbulence, though with reduced inverse energy transfer rate compared to that estimated for 2D turbulence. Recent theoretical efforts (by means of both analitical and numerical simulations) focused on the study of the inverse cascade process in 3-dimensional turbulent flows \citep[e.g.]{bif12,dub13}. \citet{bif12} presented an exact decomposition of the Navier-Stokes equation and showed that triadic interactions between waves with equally signed helicity result inverse cascade of energy, with a $-5/3$ slope as well. 

Let us now consider then that the inverse cascade operates in the shocked gas of the ISM. 
Under the assumption that the inverse cascade operates at constant 
energy transfer rate $\epsilon$, the turbulent spectrum driven at 
small scales will peak at different scales as a function with time, given as $l_{\rm peak} 
\sim \epsilon^{1/2} t^{2/3}$. Transonic perturbations driven at shock lengthscales (i.e. $\sim$pc scales, 
with $\epsilon \sim 10^{-4}$cm$^2$s$^{-3}$) would have their spectrum shifted towards 
larger scales, say $L =100$pc, at $t > 20$Myrs. With respect to the turbulent amplitude,
if dissipation is neglected, the turbulent specific energy (erg g$^{-1}$) 
grows linearly with time, i.e. $\left< v \right> \simeq 
\left( \epsilon t \right)^{1/2}$. For a constant transonic driving, e.g. at parsec scale, 
one obtains an averaged turbulent amplitude $> 10$km s$^{-1}$, at $t > 10$Myrs. Both timescales are larger than that expected for the interaction between the gas and the spiral arm. 
Still, the amplitude of turbulence induced by NTLI grown at local shocks by the  is hardly expected to be supersonic. In a more realistic scenario, the dissipation of supersonic flows is likely to dominate over the slow inverse cascade and, even if energy could diffuse towards large scales, it would result in transonic/subsonic turbulence \citep{folini2014}, in contrast to the observations \citep[e.g.][]{lar81,arm95,hey12,poidevin13}.

From our simulations turbulence is supersonic at large scales since very early stages of the run. If this is compared to the timescales needed for the inverse cascade to operate, it is clear that the shock induced turbulence is not the main turbulent driver in these models. Also, the spectra of vorticity in Fig.\ref{fig:spectra} revealed that the power at the smallest scales, equivalent to the shock thickness, rise with time, consistent to a driving at larger scales. 

Therefore, if the local converging flows (gas-arm shock) are not the main source of energy for the observed turbulence, we now must determine the role of the other main source of energy in the system: the gravitational potential of the arm.

\subsection{Driving Turbulence at Large Scales: a toy model}

If subject to an uniform gravitational field any interstellar cloud would be homogeneously accelerated, 
and internal turbulence would not be driven (at least not due to the gravitational potential). 
In a more realistic gravitational field however an extended cloud would be distorted by tidal effects. 

As described previously, we made use of a cylindrically symmetric gravitational potential 
for the spiral arm with an amplitude that radially decreases with the radial distance to 
its axis of symmetry, $r$. Let us consider each portion of a gas cloud interacting with the potential 
of the arm as an independent body, a point source, and neglect all forces except gravity. The equations of 
motion, in the direction perpendicular to the axis of symmetry of the arm, is obtained by:

\begin{equation}
\frac{ \partial \mathcal{L} }{\partial r} - \frac{d}{d t}\left( \frac{\mathcal{L} }{\dot{r}}\right) = 0.
\end{equation}

\noindent
with $\mathcal{L} = \left(\dot{\bf r} + {\bf \Omega}_{\rm p} \times {\bf r} \right)^2/2+\Phi$, the Lagrangian per unit of mass of the system in the reference frame of the spiral pattern.
The gravitational potential of the arm $\Phi$ is given by Eq.\ref{eq:potencial}. 
Let us first consider whether the arm's potential is 
dominant over the non-inertial terms of the Lagrangian. 
For each individual parcel of the gas interacting with the spiral 
arm the equation of motion perpendicular to the arm, under the dominant arm approximation, is: 

\begin{equation}
\ddot{r} \simeq -C(R_0) r \exp{\left[-\frac{r}{\sigma_\perp^2}\right]},
\label{eq:motion}
\end{equation}

\noindent
where $\sigma_\perp = \sigma \sin{i}$, $C(R_0)=\xi_0 R_0 \sigma_\perp^{-2}$, and for which we used the coordinate change $r = (R \cos{\theta-\alpha}-R_0)\cos{i}$, $r$ the coordinate of the fluid parcel perpendicular to the arm, $R$ and $R_0$ the galactocentric radii of the particle and the reference frame, respectively, 
and the fact that $r \ll R_0$. 
The right hand side of Eq.\ref{eq:motion} is time-independent, so we can use $\ddot{r} = \dot{r} \partial_{r} \dot{r}$ to obtain the quite obvious conservation equation given below:

\begin{equation}
v^2(r) \simeq v^2(r_{0})+ C(R_0) \sigma_\perp^2 \left[ \exp{\left(-\frac{r^2}{\sigma_\perp^2}\right)} - 
\exp{\left(-\frac{r_0^2}{\sigma_\perp^2}\right)}\right],
\label{eq:vperp}
\end{equation}

\noindent
where $v$ represents the linear velocity of the fluid in the local reference frame. The equation above, for an initial condition $v^2(r_0) \rightarrow 0$, and $r \ll r_0$, being $r_0$ the initial distance of the fluid parcel with respect to the arm, results in:

\begin{equation}
r(t) \simeq \sigma_\perp {\rm erfi}^{-1}\left(\frac{2\sigma_\perp C(R_0)^{1/2} t}{\sqrt{\pi}} + B(r_0)\right),
\label{eq:rperp}
\end{equation}

\noindent
being $B(r_0)$ the integration constant for $r_0$, and ${\rm erfi}^{-1}$ the inverse of the imaginary error function, 
which is expanded as a series of polinomials:

\begin{equation}
\sum_{k=0}^{\infty} \frac{(r(t)/\sigma_\perp)^{2k+1}}{k!(2k+1)} \simeq \frac{2\sigma_\perp C(R_0)^{1/2} t}{\sqrt{\pi}}+B(r_0),
\label{eq:rperpexp}
\end{equation}

Now considering an ensemble of particles, initially at rest, emerging from a region of size $L$, 
with baricenter located at $r=r_0$, i.e from a region $[r_0-L/2,r_0+L/2]$ away of the arm. 
This would mimic, in 1-dimension, a cloud of gas falling in the spiral arm. The average dispersion 
of velocities may then be estimated by computing the relative velocity of 
each element with respect to each other. At a given time $t$, each element would be located at 
positions $r$ and $r'$, being the squared relative velocity defined as $\delta v^2 \equiv 
[v(r)-v(r')]^2$. In the first order approximation\footnote{Notice that truncating the expansion (at first 
order only) is a good approximation for the case in study. If we want to study the turbulence at the densest 
regions of the spiral arms then $r<\sigma_\perp$. For $r=0.5\sigma_\perp$, for instance, we obtain 
$\mathcal{O}(1)\simeq 12\mathcal{O}(3)$. Therefore, for regions as close to the arm axis as $r=0.5\sigma_\perp$, 
we can neglect the contribution from all terms higher than the first order.}, the average velocity dispersion 
is obtained by integrating $\delta v^2$ over all the positions of all elements, as 
given below:

\begin{eqnarray}
<\delta v^2> = f \int_{-\frac{L}{2}}^{\frac{L}{2}} [v(r)-v(r')]^2 dl \nonumber \\
\simeq C(R_0) \sigma_\perp^2\int_{-\frac{L}{2}}^{\frac{L}{2}} \{ \exp\left[{-\frac{(\sigma_\perp C(R_0)^{1/2}t-r_0)^2}{2\sigma_\perp^2}}\right]- 
\nonumber \\ \exp\left[{-\frac{(\sigma_\perp C(R_0)^{1/2}t-(r_0-l))^2}{2\sigma_\perp^2}}\right]\}^2 dl 
\end{eqnarray}

\noindent
where $f$ is the normalization factor given the integral over all elements of the cloud. 

\begin{eqnarray}
<\delta v^2>  \simeq C(R_0) \sigma_\perp^3 \times \{  
\exp\left[{-\frac{(A(t)-r_0)^2}{\sigma_\perp^2}}\right]+\nonumber \\
\frac{\sqrt{\pi}}{2}\left[{\rm erf}\left(\frac{A(t) - r_0 + L/2}{\sigma_\perp}\right)- {\rm erf}\left(\frac{A(t) - r_0 - L/2}{\sigma_\perp}\right) \right]-\nonumber \\
\sqrt{\frac{\pi}{2}}\exp\left[{-\frac{(A(t)-r_0)^2}{\sigma_\perp^2}}\right] \times \nonumber \\
 \left[ {\rm erf}\left(\frac{A(t) - r_0 + L/2}{\sqrt{2}\sigma_\perp} \right) - 
{\rm erf}\left(\frac{A(t) - r_0 - L/2}{\sqrt{2}\sigma_\perp} \right) \right] \}
\label{eq:dif}
\end{eqnarray}

\noindent
where $A(t)=\sigma_\perp C(R_0)^{1/2}t$. If $L \ll 2r_0$, Eq.\ref{eq:dif} resumes to:

\begin{eqnarray}
<\delta v^2>  \sim C(R_0) \sigma_\perp^3 \frac{\sqrt{\pi}}{2} \times \nonumber 
\\ \left[{\rm erf}\left(\frac{A(t) - r_0 + L/2}{\sigma_\perp}\right)- {\rm erf}\left(\frac{A(t) - 
r_0 - L/2}{\sigma_\perp}\right) \right] ,
\end{eqnarray}

\noindent
or

\begin{eqnarray}
\frac{<\delta v^2>}{v^2(0)} \sim \sigma_\perp \frac{\sqrt{\pi}}{2} \left(1-\exp{\frac{-r_0^2}{\sigma_\perp^2}} \right)\times \nonumber 
\\ \left[{\rm erf}\left(\frac{A(t) - r_0 + L/2}{\sigma_\perp}\right)- {\rm erf}\left(\frac{A(t) - 
r_0 - L/2}{\sigma_\perp}\right) \right].
\label{eq:disp}
\end{eqnarray}

\begin{figure}
{\centering
 \includegraphics[width=7.0cm]{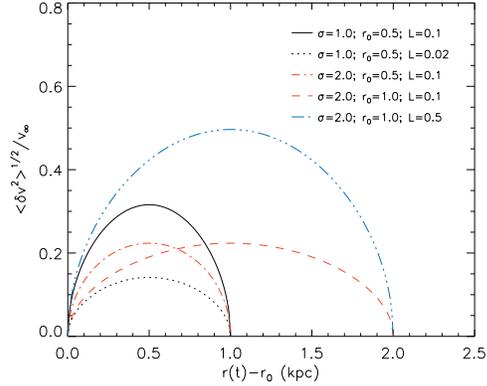}
}
 \caption{Normalized average dispersion of velocity ($\left\langle \delta v^2 \right\rangle^{1/2}/v(0)$) within the cloud, as a function of $r$ - the perpendicular distance to the spiral pattern -, obtained from Eq.\ \ref{eq:disp}. The dispersion is understood as the averaged relative velocity of the different elements of the cloud as it tidally interacts with the arm. The width of the spiral arm $\sigma$, the initial distance of the cloud baricenter to the axis of symmetry of the arm $r_0$, and the initial size of the cloud $L$, are given in kiloparsecs. Typical ISM clouds interacting with the spiral arm of our Galaxy give rise to average dispersion of velocities of $<\delta v^2>^{1/2} \sim 0.2 - 0.5 v(0)$. The parameter $v(0)$ depends on the galactocentric radius $R_0$ and the mass density of the arm. For $\xi_0 \sim 600$km$^2$s$^{-1}$kpc$^{-1}$ and 
$R_0=8$kpc, one obtains $<\delta v^2>^{1/2} \sim 30 - 46$km s$^{-1}$.}
\label{fig:dispanal}
\end{figure}

The solution for Eq.\ \ref{eq:disp} is given for a set of parameters $\sigma_\perp$, $r_0$ and $L$, 
in Figure \ref{fig:dispanal}, as a function of the displacement of the cloud baricenter ($r(t)-r_0$). 
As long as $\sigma_\perp > r_0 > L$, as discussed above, the values shown represent a 
reasonable approximation. For clouds as large as 100pc, and 
$\sigma_\perp \simeq 2r_0 \sim 1$kpc, the potential well of the arm drives a velocity 
shear that is roughly $\sim 20\% - 30\%$ of the bulk velocity of the cloud. For $\xi_0 \sim 600$km$^2$s$^{-1}$kpc$^{-1}$ and 
$R_0=8$kpc, one obtains $<\delta v^2>^{1/2} \sim 30 - 46$km s$^{-1}$, at the lengthscale of the cloud size, i.e. 
few tens to a hundred of parsecs.

We must point out that Eq.\ \ref{eq:disp} gives an estimate for the internal shear of the cloud, not the 
turbulence itself. The transfer of this local kinetic energy into turbulence will depend on the 
processes taking place during the passage of the cloud through the arm. For instance if
 the cloud is retained in the arm for a long time, at least one dynamical timescale, the efficiency is 
high.

\section{Discussion}

SNe are among the main energy sources in the ISM. Naturally, one would consider its contribution in 
driving the turbulence amplitudes ($>10$km s$^{-1}$) to be dominant given the absence of other 
efficient feedback mechanisms. Despite of the energy input, SNe driven models strugle to explain 
other features of the ISM turbulence, such as its universality and the observed scalings. The ISM turbulence is universal, and shows no correlation to the local star formation rate. Also, observations reveal that 
turbulence may be driven at scales larger than 100pc. Such driving scales would only be reasonably explained in a SNe driven model by superbubbles, i.e. many SNe working together to form large scale strucutures. Again universality issues apply. Other issues have to be addressed as well, such as the shielding of dense and cold structures to the motions of the diffuse medium, as quiescent molecular clouds present turbulent motions that could not be triggered by an external source (at least not by ram pressures).

In the work presented here we focused on the processes that may trigger turbulence at large scale in the ISM. 
We showed that the tidal interaction with the gravitational potential of an spiral arm is 
responsible for driving complex internal motions in an interstellar cloud. Notice that the interaction between the arm and the cloud is not due to a shock, as in a converging flows approach, but by its tidal differential acceleration. Converging flows - or shocks of ISM gas with spiral arms - are an interesting mechanism that can explain the formation of molecular clouds in the ISM, as discussed in Section 1, but fail in properly feed turbulence as it is observed. An advantage of these models is its universality in spiral galaxies. As pointed before, radiative shocks in converging flows \citep[see][]{hei2005, aud05, hei06, bonnell06,bonnell13} have been shown to be very efficient in forming dense and cold structures in the ISM, but not very efficient in driving supersonic turbulence though. Theoretical studies of these systems reveal that the main process that lead to the structuring of clumps is the NTLI, which is also known to result in small scale subsonic/transonic turbulent flows \citep{vish94,
folini2014}. 

The present model provides a different view of the same problem: the gravitational interaction between an inhomogeneous ISM and the spiral pattern of the Galaxy. Inhomogeneities in the ISM naturally arise at the spiral arms due to shocks, cooling and/or gravitational fragmentation, which move to the interarm region eventually \citep[see][]{bonnell13}. Such structures then interact with the next spiral arm, as described in this work.
The interaction is not only collisional, but gravitational. Large ISM inhomogeneities, such as $>100$pc scale neutral clouds, would suffer differential accelerations that drive internal motions. The internal shear develops into turbulence at scales as large as the cloud size.

\citet{hei06}, for instance, compared the kinetic energy of the inflow to that in
the unstable and cold gas phase, obtaining turbulence driving efficiencies of order of $2-5$\%. The efficiency in our models 
may be obtained by comparing the turbulent kinetic energy to the gravitational potential variation during 
the crossing. In the models were the clouds are trapped to the arm, the efficiency is $>25$\%.
This difference occurs basically because the turbulence in converging flow models is a post-product of 
the radiative shocks, i.e. firstly kinetic energy has to be lost efficiently for the formation of the 
cold and dense layer, which is then non-uniformly accelerated by the NTLI. Naturally, most of the initial reservoir of energy is lost by radiation at the shock. Instead, the mechanism reported here benefits directly from the large scale kinetic energy due to the flow into the spiral arms gravitational potential.

We must point out here that both processes actually occur in our simulations. The importance of each can be 
determined as detailed earlier. We analyzed the turbulent motions generated in the clouds as a function of different 
parameters of the simulation, and found no correlation of the driven turbulence with the relative velocity between the cloud and the arm. The spiral pattern parameters, however, show a clear correlation with the driven turbulence. Larger $\xi_0$ values 
result in larger turbulent amplitudes, while larger $\sigma_\perp$ result in smaller $<\delta v^2>^{1/2}$. 
These two correlations support the idea of a tidally driven turbulence for clouds in the ISM. More massive 
arms present stronger differential accelerations within the cloud, which leads to larger internal shearing. 
More compact arms are responsible for larger turbulent amplitudes as well. 
Notice that the interesting parameter is not $\sigma_\perp$ itself but $r_0/\sigma_\perp$, i.e. the position of the cloud at the potential profile. If the cloud is 
positioned in a region that $r_0/\sigma_\perp<1$, the net result of the interaction with the arm will be reduced. 

Our model is different to that of turbulence driven globally in the galaxy by gravitational instability \citep{wada02}. In the later, the collapse of gravitationally untable interstellar gas is responsible for driving the observer dispersion of velocity. In our model the gravitational interaction between the gas and the potential of the arms would be responsible for sheared motions that further evolve into turbulence. The main difference between both is that the gravitational collapse drives motions at small scales ($\sim \lambda_{\rm Jeans}$), with subsonic/transonic motions \citep{age09}, while the gas-arm interaction drives supersonic turbulence at large scales. Also, the gravitational collapse drives coherent inward motions that may develop into chaotic motions after the complex interaction between collapsed structures. This is possibly the cause of the flatter power spectra observed in simulations of gravitationally collapsing disks. It is particularly interesting to perform more numerical simulations of the gas-arm interaction, as done in the present work, but considering self-gravity to account for the possible effects of gravitational collapse in the turbulence driving. This will be pursued in a future work.

From an observational perpective there are recent high spatial resolution data available for few nearby spiral galaxies, for which a detailed study of the ISM turbulence have been provided \citep[see][]{hugh13a, hugh13b, colo14}. \citet{hugh13b} and \citet{colo14} showed that cloud-scale CO linewidths are typically related to the arm/interarm properties, as predicted in our model. Also, \citet{hugh13a} presented observed $^{12}$CO(1-0) line profiles for different regions of M51 showing that the linewidths are larger at regions with higher stellar surface densities, indicating that regions of deepest potential are indeed more turbulent.

It is worth mentioning here that the mechanism of turbulent driving by tidal forces acting on the ISM inhomogeneities is maximized 
if the cloud interacts with the arm for longer timescales. As the cloud bounces in the potential well the gravitational energy of the cloud is effectively drained into turbulent motions. The process known as cloud trapping, or cloud 
streaming, has already been discussed previously, mostly in the context of crowding stellar orbits in spiral galaxies, and on the formation of Giant Molecular Clouds (GMCs).

\citet{rs87} for instance, studied the orbital dynamics of clouds and stars in N-body simulations, in which angular momentum losses due to cloud-cloud collisions were taken into account. The authors showed that streaming of clouds along the arm, i.e. orbits partially trapped by the spiral pattern, was present even without cloud-cloud collisional dissipation. The streaming (trapping) naturally arises due to the tendency for orbits to crowd at the spiral arms. However, as the clouds stream towards smaller radii they are accelerated and eventually leave the arm. These authors found typical timescales for the crowding as $\simeq 50$Myrs, and that trapping is enhanced for even longer timescales if consideable linear/angular momentum is lost by the cloud. Unfortunately, these authors were not able to distinguish internal motions within the clouds due to numerical limitations. It is, however, interesting to compare these results to our own, which makes use of different dissipation mechanisms. Such radial inwards/outwards 
streaming flows, related to the spiral patterns, have been observed in numerical simulations \citep[e.g.][]{dobbs06,she07}, and observationally \citep[e.g.][]{aalto99,fre05,rif08,mei13}.  The cloud trapping is particularly important to the fact that clouds would rarely interact with arms more than once in a dynamical timescale. 

As the clouds are pertubed by the arms we expect the clouds to fragment and collapse, or to be dissipated as the Coriolis and centrifugal effects result in its migration to the interarm regions. Eventually, if the cloud survives as individual entity, we expect the internal turbulence to decay quickly, in a timescale of $\tau \sim L/v_L$, compared to the time between subsequent arm crossings, $\delta t_{\rm arms} \sim 2 \pi R/m\left|V(R) - \Omega_p R \right|$. It is unprobable then that the turbulence in interstellar clouds to be built up with time as a consequence of many interactions with different arms. Therefore we believe that the maximum kinetic energy provided by this mechanism is limited by that of one arm crossing.

Notice that the typical streaming process that natually arise from the torques of the spiral pattern on point sources is not a dissipative mechanism. For the purposes of keeping the cloud close to the arm for long timescales this is not important anyway. Related dissipative models have been proposed also \citep[e.g.][]{zhang96}. In the tidal model presented here energy is ``lost" as the cloud interacts with the potential of the arm, as well. Most of the ``loss" is not due to internal friction (i.e. heat) but to the conversion of the large scale motions in the cloud into smaller scales, due to a kinetic cascade. Eventually part of this energy is dissipated into heat, while the rest remains as randomized kinetic energy of the dense structures formed in the process. This energy loss can be estimated from Eq.\ref{eq:disp}, as shown in Fig.\ref{fig:dispanal}. In one cloud-arm passage the internal dissipation of the potential energy may be as large as 25\% of the escape kinetic energy.  
Such a process can therefore enhance the timescales by which the clouds interact with arms, as well as, possibly result in increased radial motions of gas in spiral galaxies. 

These results have been presented in the scenario where arms form as long-lived perturbations in the gravitational potential of the disk. Let us now consider a different context, in which the spiral arms are transient \citep{baba13} and giant HI/molecular clouds are formed by gravitational collapse of the insterstellar gas, instead of by standing shocks with spiral arms \citep[see][]{wada02,dobbs14}. In such a scenario multiple and discontinous arms are formed - and destroyed - in relatively short timescales. Also, dense and cold regions would form, scarcely distributed in the disk, as a consequence of the gravitational collapse of a cooling ISM \citep{age09}. The relative motion between the clouds and the arms would be reduced compared to the standing shock wave model. Such a scenario has been related to Sa-type spiral galaxies, while prominent standing waves in stellar motions would correspond better to types Sb and Sc. One then may wonder what happens to the turbulence driving mechanism proposed in this work in a transient arm scenario.

Despite the dynamical differences between the two galactic scenarios the driving mechanism occurs similarly in both. In Section 4.3 the driving observed in the simulations was described as the consequence of the tidal interaction between the potential well of the arm and an interstellar cloud, and the non-linear evolution of internal sheared motions. We showed that the relative motion between them is not an initial condition\footnote{The condition $v(r_0) \rightarrow 0$ has even been assumed for Eq.\ 16.}(see Eq.\ 16), and the turbulent amplitude is basically dependent on the gravitational potential, and width, of the arm, and on the size of the cloud. Therefore we do not expect any difference in the turbulent amplitude in the case of a transient spiral arm scenario, given the timescales for cloud-arm interaction are short compared to the dissipation timescale of the arms.

\section{Conclusions}

In this work we studied the onset of turbulence in the ISM based on the interaction of 
interstellar gas inhmogeneities and the spiral arm. Here we focused only on the gravitational 
interaction of the spiral pattern with the ISM. 3-dimensional hydrodynamical 
simulations are provided with different initial setups. In all models turbulence is observed, 
at different locations and levels. In most of the models the cloud interacting 
with the arm becomes strongly turbulent $\left<\delta v^2\right> \gg c_s$. In contrast to previous 
theories to account for the ISM turbulence, the injection here occurs at 
large scales and is not related to local properties such as star formation rates. 
The statistics of turbulence obtained for the models are in agreement with a Kolmogorov type 
turbulent cascade, and synthetic observables are compatible to the Larson scaling relations.

Although our simulations span only a limited intertial range, we can draw conclusions that should apply to the full range of turbulence seen in molecular clouds.

We find that the spiral shock can trigger turbulence, but in contrast to pure colliding flows, it is due to the large scale tidal interaction rather than the small scale of the shock-induced fluid instabilities.
Naturally, the differential 
forces act at large scales ($l_{\rm max} \simeq L_{\rm cloud}$). The sheared motions within the cloud 
then develop Kelvin-Helmholtz, as well as internal shocks, which evolve into a turbulent cascade 
later on. From the first cloud-arm crossing, typical timescales of $20 - 50$Myrs are required for the 
turbulence to develop. 

More massive and more compact arms, i.e. larger $\xi_0$ and smaller $\sigma_\perp$, respectively, result 
in larger turbulent amplitudes ($<\delta v^2>$). No correlation has been obtained between $<\delta v^2>$ 
and the galactocentric radius of the cloud - by means of the cloud-arm relative velocity. Therefore, 
turbulence would be ``universal", at least near the spiral pattern of the Galaxy. Though our results have been addresses in a scenario of long-lived standing arms, i.e. based on the density wave theory, these should be similar in transient arm scenarios, given that clouds interact with a non-uniform stellar potential well. 

An analytical toy-model is presented to account for the random motions generated within a cloud interacting 
with a spiral arm. The analytical model predictions are in agreement with the main results of the simulations, 
confirming that the main triggering mechanism of the observed turbulence is the differential gravitational forces within the cloud. These results are in agreement with recent observations with high spatial resolution of nearby spiral galaxies \citep[e.g.][]{hugh13a,hugh13b,colo14}.

The models were performed without self-gravity. A natural consequence of this work would be to study next 
the effects of self-gravity in such a model, where fragmentation and collapse of small structures would be allowed. This is to be studied in a future work.

\section*{Acknowledgments}

The authors thank the anonymous referee for the valuable comments provided.
DFG thanks the European Research Council (ADG-2011 ECOGAL), and
Brazilian agencies CAPES (3400-13-1)
and FAPESP (no.2011/12909-8) for financial support. IB ackowledges the 
European Research Council (ADG-2011 ECOGAL) for financial support. 
GK acknowledges support from FAPESP (grants no. 2013/04073-2 and 2013/18815-0).


\begin{thebibliography}{}

\bibitem[Aalto et al.(1999)]{aalto99} Aalto, S., 
H{\"u}ttemeister, S., Scoville, N.~Z., 
\& Thaddeus, P.\ 1999, ApJ, 522, 165

\bibitem[Agertz et al.(2009)]{age09} Agertz, O.; Lake, G.; Teyssier, R.; Moore, B.; Mayer, L. \& Romeo, A. 2009, MNRAS, 392, 294

\bibitem[Antoja et al.(2011)]{ant11}Antoja, T., Figueras, F., Romero-G\'omez, M., et al. 2011, MNRAS, 418, 1423

\bibitem[Armstrong, Rickett \& Spangler(1995)]{arm95} Armstrong, J. W., Rickett, B. J., \& Spangler, S. R. 1995, ApJ, 443, 209

\bibitem[Audit \& Hennebelle(2005)]{aud05} Audit E. \& Hennebelle P., 2005, A\&A, 433, 1

\bibitem[Baba et al.(2009)]{baba09} Baba, J., Asaki, Y., Makino, J., et al.\ 2009, ApJ, 706, 471

\bibitem[Baba, Saitoh \& Wada(2013)]{baba13} Baba, J., Saitoh, T. R. \& Wada, K. 2013, ApJ, 763, 46

\bibitem[Ballesteros-Paredes et al.(2011)]{bal11} 
Ballesteros-Paredes, J., Hartmann, L.~W., V{\'a}zquez-Semadeni, E., 
Heitsch, F., \& Zamora-Avil{\'e}s, M.~A.\ 2011, MNRAS, 411, 65 

\bibitem[Banerjee et al.(2009)]{ban09} Banerjee R., V\'azquez-Semadeni E., Hennebelle P. \& Klessen R. S., 2009,
MNRAS, 398, 1082


\bibitem[Biferale et al.(2012)]{bif12} Biferale, L., 
Musacchio, S., \& Toschi, F.\ 2012, Physical Review Letters, 108, 164501 

\bibitem[Boffetta 
\& Ecke(2012)]{bof12} Boffetta, G., \& Ecke, R.~E.\ 2012, Annual Review of Fluid Mechanics, 44, 427 

\bibitem[Bonnell et al.(2006)]{bonnell06} Bonnell, I. A., Dobbs,  C. L., Robitaille, T. P. \& Pringle, J. E., 2006, MNRAS, 365, 37.

\bibitem[Bonnell et al.(2013)]{bonnell13} Bonnell, I.~A., Dobbs, 
C.~L., \& Smith, R.~J.\ 2013, MNRAS, 430, 1790

\bibitem[Bottema(2003)]{bott03} Bottema, R.\ 2003, MNRAS, 344, 358

\bibitem[Carlberg \& Freedman(1985)]{carl85} Carlberg, R.~G., \& Freedman, W.~L.\ 1985, ApJ, 298, 486

\bibitem[Colombo et al.(2014)]{colo14} Colombo, D., Hughes, 
A., Schinnerer, E., et al.\ 2014, ApJ, 784, 3

\bibitem[Dobbs \& Bonnell(2006)]{dobbs06} Dobbs, C. L. \& Bonnell, I. 2006, MNRAS, 367, 873

\bibitem[Dobbs et al.(2008)]{dobbs08} Dobbs, C. L.; Glover, S. C. O.; Clark, P. C.; Klessen, R. S. 2008, MNRAS, 389, 1097

\bibitem[Dobbs \& Baba(2014)]{dobbs14} Dobbs, C. L.; Baba, J. 2014, PASA, in press (arxiv:1407.5062)

\bibitem[Dubief et al.(2013)]{dub13} Dubief, Y., Terrapon, 
V.~E., \& Soria, J.\ 2013, Physics of Fluids, 25, 110817 

\bibitem[Elmegreen \& Thomasson(1993)]{elme93} Elmegreen, B.~G., \& Thomasson, M.\ 1993, A\&A, 272, 37

\bibitem[Elmegreen \& Scalo(2004)]{elm04} Elmegreen, B. \& Scalo, J. 2004, ARA\&A, 42, 211

\bibitem[Falceta-Gon\c calves et al.(2010a)]{fal10a} Falceta-Gon\c calves, D., de Gouveia Dal Pino, E. M., Gallagher, J. S., \&  Lazarian, A., 2010, ApJ, 708, L57

\bibitem[Falceta-Gon\c calves et al.(2010b)]{fal10b} Falceta-Gon\c calves, D., Caproni, A., Abraham, Z., Teixeira, D. M., \& de Gouveia Dal Pino, E. M., 2010, ApJ, 713, L74

\bibitem[Falceta-Gon\c calves et al.(2010c)]{fal10c} Falceta-Gon\c calves, D., Lazarian, A., \& Houde, M., 2010, ApJ, 713, 1376

\bibitem[Falceta-Gon\c calves \& Lazarian(2011)]{fal11} Falceta-Gon\c calves, D. \& Lazarian, A., 2011, ApJ, 735, 99

\bibitem[Falceta-Gon{\c c}alves \& Abraham(2012)]{fal12} Falceta-Gon\c calves, D., \& Abraham, Z.\ 2012, MNRAS, 423, 1562

\bibitem[Falceta-Gon{\c c}alves 
\& Monteiro(2014)]{fm14} Falceta-Gon{\c c}alves, D., \& Monteiro, H.\ 2014, MNRAS, 438, 2853 

\bibitem[Falceta-Gon{\c c}alves et al.(2014)]{fal14} Falceta-Gon{\c c}alves, D., Kowal, G., Falgarone, E. \& Chian, A.-L.\ 2014, Nonlinear Processes in Geophysics, 21, 587

\bibitem[Fresneau et al.(2005)]{fre05} Fresneau, A., Vaughan, 
A.~E., \& Argyle, R.~W.\ 2005, AJ, 130, 2701

\bibitem[Folini et 
al.(2014)]{folini2014} Folini, D., Walder, R., \& Favre, J.M.\ 2014, A\&A, 562, A112

\bibitem[Fujii et al.(2011)]{fujii11} Fujii, M. S., Baba, J., Saitoh, T. R., Makino, J., Kokubo, E., \& Wada, K.
2011, ApJ, 730, 109

\bibitem[Gerhard(2011)]{ger11} Gerhard O. 2011, Mem. Soc. Astron. Ital. Suppl. 18, 185

\bibitem[Goldsmith et al.(2008)]{gol08} Goldsmith, P. F.; Heyer, M.; Narayanan, G.; Snell, R.; Li, D.; Brunt, C.: 2008, ApJ, 680, 428

\bibitem[Gressel et al.(2008)]{gres08} Gressel, O.; Elstner, D.; Ziegler, U.; Rudiger, G., 2008, A\&A, 486, 35

\bibitem[He et al.(2011)]{he11}
He, Z., Li, X., Fu, D., \& Ma, Y.\ 2011, ScChG, 54, 511

\bibitem[Heitsch et al.(2005)]{hei2005} Heitsch, F., Burkert, 
A., Hartmann, L.W., Slyz, A.D., 
\& Devriendt, J.E.G.\ 2005, ApJL, 633, L113

\bibitem[Heitsch et al.(2006)]{hei06} Heitsch, F.; Slyz, A. D.; Devriendt, J. E. G.; Hartmann, L. W.; Burkert, A. 2006, ApJ, 648, 1052

\bibitem[Heitsch, Naab \& Walch(2011)]{hei11} Heitsch, F.; Naab, T.; Walch, S. 2011, MNRAS, 415, 271

\bibitem[Henley et al.(2010)]{hen10} Henley, D.B.; Shelton, R.L.; Kwak, K; Joung, M.R. \& Mac Low, M.-M. 2010, ApJ, 723, 935

\bibitem[Hennebelle et al.(2007)]{hen07} Hennebelle P., Audit E. \& Miville-Desch\^enes M.-A., 2007, A\&A, 465, 445

\bibitem[Hennebelle \& Falgarone(2012)]{hen12} Hennebelle, P. \& Falgarone, E. 2012, ARA\&A, 20, 55

\bibitem[Heyer \& Brunt(2004)]{hey04} Heyer, M.H. \& Brunt, C.M. 2004, ApJ, 615, L45

\bibitem[Heyer et al.(2009)]{hey09} Heyer, M., Krawczyk, C., Duval, J., \& Jackson, J.M. 2009, ApJ, 699, 1092

\bibitem[Heyer \& Brunt(2012)]{hey12} Heyer, M.H. \& Brunt, C.M. 2012, MNRAS, 420, 1562

\bibitem[Hill et al.(2012)]{hill12} Hill, A. S.; Joung, M. R.; Mac Low, M-M.; Benjamin, R. A.; Haffner, L. M.; Klingenberg, C.; Waagan, K. 2012, ApJ, 750, 104

\bibitem[Hughes et al.(2013a)]{hugh13a} Hughes, A., Meidt, 
S.~E., Schinnerer, E., et al.\ 2013, ApJ, 779, 44

\bibitem[Hughes et al.(2013b)]{hugh13b} Hughes, A., Meidt, 
S.~E., Colombo, D., et al.\ 2013, ApJ, 779, 46

\bibitem[Hunter et al.(1986)]{hunter84} Hunter, J.H., Jr., Sandford, M.T., II, Whitaker, R.W., \& Klein, R.I.\ 1986, ApJ, 305, 309 

\bibitem[Inoue \& Fukui(2013)]{inoue2013} Inoue, T., \& Fukui, Y.\ 2013, ApJL, 774, L31

\bibitem[Junqueira et al.(2013)]{jun13} Junqueira,T. C.; L\'epine, J. R. D., Braga, C. A. S. \&  Barros, D. A. 2013, A\&A, 550, A91

\bibitem[Kalnajs(1973)]{kalnajs73} Kalnajs A. J., 1973, PASA, 2, 174

\bibitem[Kim \& Ostriker(2006)]{kim06} Kim, W.-T. \& Ostriker, E. C. 2006, ApJ, 646, 213

\bibitem[Kim, Kim \& Ostriker(2006)]{kimkim06} Kim, C.-G.; Kim, W.-T.; Ostriker, E. C. 2006, ApJ, 649, 13L

\bibitem[Kim, Kim \& Kim(2014)]{kkk14} Kim, W.-T.; Kim, Y.; Kim, J.-G. 2014, ApJ, 789, 68

\bibitem[Kowal \& Lazarian(2010)]{kowal10} Kowal, G. \& Lazarian, A., 2010, ApJ, 720, 742

\bibitem[Kowal et al.(2011a)]{kowal11a} Kowal, G., de Gouveia Dal Pino, E. M., \& Lazarian, A., 2011, ApJ, 735, 102

\bibitem[Kowal et al.(2011b)]{kowal11b} Kowal, G., Falceta-Gon\c calves, D. A., \& Lazarian, A. 2011, NJPh, 13, 3001

\bibitem[Kraichnan(1967)]{kra67} Kraichnan, R.H.\ 1967, Physics of Fluids, 10, 1417 

\bibitem[Larson(1981)]{lar81} Larson, R.B. 1981, MNRAS, 194, 809

\bibitem[L\'epine et al.(2008)]{lep08} L\'epine, J. R. D., Dias, W. S., \& Mishurov, Y. N. 2008, MNRAS, 386, 2081

\bibitem[Lin \& Shu(1964)]{lin64} Lin, C. C. \& Shu, F. H. 1964, ApJ, 140, 646

\bibitem[Liu, Wu \& Zhang(2012)]{liu12} Liu T., Wu Y. \& Zhang H. 2012, ApJS, 202, 4

\bibitem[Mac Low \& Klessen(2004)]{mac04} Mac Low, M.M \& Klessen, R. S. 2004, Reviews of Modern Phys., 76, 125

\bibitem[Meidt et al.(2013)]{mei13} Meidt, S.~E., Schinnerer, 
E., Garc{\'{\i}}a-Burillo, S., et al.\ 2013, ApJ, 779, 45

\bibitem[Melioli et al.(2009)]{melioli09} Melioli, C.; Brighenti, F.; D'Ercole, A.; de Gouveia Dal Pino, E. M. 2009, MNRAS, 399, 1089

\bibitem[Mignone \& Bodo(2006)]{mignone06} Mignone, A., \& Bodo, G.\ 2006, MNRAS, 368, 1040

\bibitem[Minter \& Spangler(1996)]{min96} Minter, A.H. \& Spangler, S.R. 1996, ApJ, 458, 194

\bibitem[Paret \& Tabeling(1998)]{paret98} Paret, J., \& Tabeling, P.\ 1998, Physics of Fluids, 10, 3126 

\bibitem[Pichardo et al.(2003)]{pich03} Pichardo, B., Martos, M., Moreno, E., \& Espresate, J. 2003, ApJ, 582, 230

\bibitem[Poidevin et al.(2013)]{poidevin13} Poidevin, F., Falceta-Gon\c calves, D., Kowal, G., de Gouveia Dal Pino, E., Magalhaes, A. M. 2013, ApJ, 777, 112

\bibitem[Roberts(1969)]{rob69} Roberts, W.~W.\ 1969, ApJ, 
158, 123 

\bibitem[Riffel et al.(2008)]{rif08} Riffel, R.~A., 
Storchi-Bergmann, T., Winge, C., et al.\ 2008, MNRAS, 385, 1129

\bibitem[Roberts 
\& Stewart(1987)]{rs87} Roberts, W.~W., Jr., \& Stewart, G.~R.\ 1987, ApJ, 314, 10 

\bibitem[Ruiz et al.(2013)]{ruiz13} Ruiz, L. O., Falceta-Gon\c calves, D., Lanfranchi, G. A. \& Caproni, A. 2013, MNRAS, 429, 1437

\bibitem[Ruuth(2006)]{ruuth06} Ruuth, S.J.\ 2006, Mathematics of Computation, 75, 183

\bibitem[Santos-Lima et al.(2014)]{rei14} Santos-Lima, R., de 
Gouveia Dal Pino, E.~M., Kowal, G., et al.\ 2014, ApJ, 781, 84 

\bibitem[Scarano \& Lepine(2012)]{scar12} Scarano Jr., S. \& Lepine, J.R.D., 2012, MNRAS, 428, 625

\bibitem[Sellwood \& Carlberg(1984)]{sellwood84} Sellwood, J. A., \& Carlberg, R. G. 1984, ApJ, 282, 61

\bibitem[Sellwood(2011)]{sellwood11} Sellwood, J. A. 2011, MNRAS, 410, 1637

\bibitem[Shetty et al.(2007)]{she07} Shetty, R., Vogel, 
S.~N., Ostriker, E.~C., \& Teuben, P.~J.\ 2007, ApJ, 665, 1138 

\bibitem[Smith, Sigurdsson \& Abel(2008)]{smith08} Smith, B., Sigurdsson, S. \& Abel, T. 2008, MNRAS, 385, 1443

\bibitem[Toomre(1969)]{toomre69} Toomre, A. 1969, ApJ, 158, 899

\bibitem[Vall\'ee(2014a)]{vallee14a} Vall\'ee, J. 2014a, AJ, 148, 5

\bibitem[Vall\'ee(2014b)]{vallee14b} Vall\'ee, J. 2014b, ApJS, in press (arxiv:1409.4801)

\bibitem[V{\'a}zquez-Semadeni et al.(2006)]{vs2006} 
V{\'a}zquez-Semadeni, E., Ryu, D., Passot, T., Gonz{\'a}lez, R.F., 
\& Gazol, A.\ 2006, ApJ, 643, 245

\bibitem[V{\'a}zquez-Semadeni et al.(2007)]{vs2007} 
V{\'a}zquez-Semadeni, E., G{\'o}mez, G.C., Jappsen, A.K., et al.\ 2007, 
ApJ, 657, 870 

\bibitem[V{\'a}zquez-Semadeni et al.(2008)]{vs2008} 
V{\'a}zquez-Semadeni, E., Gonz{\'a}lez, R.~F., Ballesteros-Paredes, J., 
Gazol, A., \& Kim, J.\ 2008, MNRAS, 390, 769 

\bibitem[Vishniac(1994)]{vish94} Vishniac, E. T. 1994, ApJ, 428, 186

\bibitem[von Hoerner(1951)]{vonH51} von Hoerner, S. 1951, Zeitschrift f\"ur Astrophysik, 30, 17

\bibitem[von Weizs\"acker(1951)]{von51} von Weizs\"acker, C. F. 1951, ApJ, 114, 165

\bibitem[Wada, Meurer \& Norman(2002)]{wada02} Wada, K.; Meurer, G. \& Norman, C.A. 2002, ApJ, 577, 197

\bibitem[Wada \& Koda(2004)]{wada04} Wada, K. \& Koda, J. 2004, MNRAS, 349, 270	

\bibitem[Wada \& Norman(2007)]{wada07} Wada K., Norman C. A., 2007, ApJ, 660, 276

\bibitem[Wada, Baba \& Saitoh(2011)]{wada11} Wada, K.; Baba, J. \& Saitoh, T. R., 2011, ApJ, 735, 1

\bibitem[Walder \& Folini(2000)]{walder2000} Walder, R., \& Folini, D.\ 2000, ApSS, 274, 343

\bibitem[Williams, Blitz \& McKee(2000)]{will00} Williams, J. P.; Blitz, L.; McKee, C. F. 2000, in Protostars and Planets IV (Book - Tucson: University of Arizona Press; eds Mannings, V., Boss, A.P., Russell, S. S.), p. 97

\bibitem[Woodward(1976)]{wood76} Woodward, P. R. 1976, ApJ, 207, 484

\bibitem[Yoshida et al.(2010)]{yos10} Yoshida, A., Kitamura, Y., Shimajiri, Y., \& Kawabe, R. 2010, ApJ, 718, 1019

\bibitem[Zhang(1996)]{zhang96} Zhang, X.\ 1996, ApJ, 457, 125

\end{thebibliography}
\end{document}